\documentclass[trackchanges,twocolumn]{aastex701}

\usepackage{newtxtext,newtxmath}
\usepackage{graphicx}	
\usepackage{amsmath}	
\usepackage{amssymb}	
\usepackage{mhchem}  
\newcounter{abcd}
\usepackage{booktabs}
\usepackage{fix-cm}

\begin{document}

\title{Deuterium fractionation and CO depletion in Barnard 5} 

\author[orcid=0000-0002-1584-3620,gname=Igor,sname=Petrashkevich]{Igor Petrashkevich}
\affiliation{Research Laboratory for Astrochemistry, Ural Federal University, Mira st. 19, 620002 Yekaterinburg, Russia}
\email[show]{petrashkevich.igor@gmail.com}  

\author[orcid=0000-0001-6004-875X,gname=Anna, sname=Punanova]{Anna Punanova} 
\affiliation{Onsala Space Observatory, Chalmers University of Technology, Observatoriev\"agen 90, R\aa\"o, Onsala, Sweden}
\email{punanovaanna@gmail.com}

\author[orcid=0000-0003-1481-7911,gname=Paola,sname=Caselli]{Paola Caselli}
\affiliation{Max Planck Institute for extraterrestrial Physics, Giessenbachstrasse 1, 85748 Garching, Germany}
\email{caselli@mpe.mpg.de}

\author[orcid=0000-0002-3972-1978,gname=Jaime Eduardo,sname=Pineda]{Jaime E. Pineda}
\affiliation{Max Planck Institute for extraterrestrial Physics, Giessenbachstrasse 1, 85748 Garching, Germany}
\email{jaime.e.pineda@gmail.com}

\author[orcid=0000-0002-9148-1625,gname=Olli,sname=Sipil\"a]{Olli Sipil\"a}
\affiliation{Max Planck Institute for extraterrestrial Physics, Giessenbachstrasse 1, 85748 Garching, Germany}
\email{osipila@mpe.mpg.de}

\author[orcid=0000-0003-1684-3355,gname=Anton,sname=Vasyunin]{Anton Vasyunin}
\affiliation{Research Laboratory for Astrochemistry, Ural Federal University, Mira st. 19, 620002 Yekaterinburg, Russia}
\email{anton.vasyunin@gmail.com}


\begin{abstract}

Deuterium fractionation provides a key diagnostic of the physical and chemical evolution of prestellar and protostellar cores, where it is strongly linked to CO depletion in cold, dense gas. We present the first spatially resolved maps of deuterium fraction and CO depletion in the Barnard 5 (B5) region of the Perseus molecular cloud, covering both a starless core and the protostellar core hosting the Class 0/I source IRAS 03445+3242. Using IRAM 30~m observations of N$_2$H$^+$(1--0), N$_2$D$^+$(1--0), H$^{13}$CO$^+$(1--0), and DCO$^+$(2--1), complemented by C$^{18}$O(2--1) data, we derive column density, deuterium fraction, and CO depletion maps. We find that the deuterium fraction in both mentioned nitrogen- and carbon-bearing species increases from the protostellar to the starless core, reaching $R_D^{\rm N_2H^+}=0.43\pm0.10$ and $R_D^{\rm HCO^+}=0.09\pm0.02$ in the starless core, compared with $0.15\pm0.03$ and $0.05\pm0.01$, respectively, in the protostellar core. The CO depletion factor also rises from $4.1\pm0.1$ to $5.0\pm0.1$ across the same transition. While the embedded YSO reduces deuteration in the dense inner gas, the less dense envelope traced by HCO$^+$ is only slightly affected at our resolution. Our analysis confirms that CO freeze-out and the presence of a protostar jointly regulate deuterium chemistry in star-forming regions.

\end{abstract}



\section{Introduction} 

Low-mass stars start forming in prestellar cores, cold dense structures of molecular clouds characterized by temperatures of $\simeq$10~K and densities of $\simeq$10$^4$--10$^7$~cm$^{-3}$ \citep[][]{Ward-Thompson1999}.
At these earliest stages of star formation, the abundances of deuterated molecules, i.e., molecules where hydrogen is replaced by deuterium, are significantly higher compared to what would be expected based on the cosmic abundance ratio of deuterium and hydrogen \citep[e.g.,][]{Friesen2013,Punanova2016}. The abundance ratios of deuterated and non-deuterated species ($R{_D}$) in prestellar clouds may reach values higher than 0.01, although the cosmological D/H abundance ratio is significantly lower \citep[1.5$\times$10$^{-5}$,][]{Dalgarno1984,Linsky2006}. Due to the lower zero-point energy of deuterium compared to protium, proton-exchange reactions at low temperatures preferentially replace hydrogen with deuterium in molecules~\citep{Dalgarno1984}. This process allows deuterated molecules to accumulate, thereby increasing the deuterium fraction, $R_D$.

The deuterium fractionation process starts with the formation of deuterated trihydrogen cation (H$_2$D$^+$). It is formed in a reaction of H$_3^+$ ion with deuterated molecular hydrogen (HD) that proceeds in one direction at low temperature: H$_3^+$ + HD $\rightarrow$ H$_2$D$^+$ + H$_2$. However, H$_3^+$ is a very reactive ion, and can competitively react with other species in the gas phase, most importantly, with CO. This reaction results in the formation of the HCO$^+$ ion. The undepleted gas-phase abundance of CO is higher than that of HD. Thus, abundant CO in the gas phase reduces the efficiency of deuterium fractionation. In fact, high CO depletion observed in prestellar cores is associated with an increase in the deuterium fraction in the gas phase, as shown in both observations \citep[e.g.,][]{Caselli2002,Bacmann2003,Crapsi2005,Socci2024} and models \citep{Caselli2008,Kong2015,Sipila2015_model,Sipila2015_spin_state}. 

Next stage of protostellar development is characterized by the formation of a young stellar object (YSO). As the kinetic temperature around it rises, the deuterium fraction in molecules decreases. This is caused both by the equilibrium shift in proton-deuteron exchange reaction mentioned above, and by the release of CO molecules back to the gas phase. Thus, deuterium fraction is highly sensitive to the physical conditions in star-forming regions. It serves as an important instrument to study prestellar phase and the first stage of protostellar core, as well as to understand the chemical processes and to impose observational constraints on chemical models \citep{Caselli2012}. 

Many observational studies are focused on the deuterium fraction in different species in cold dense cores using single pointing observations \citep[e.g.,][]{Crapsi2005,Friesen2013,Punanova2016,Harju2017,Ambrose2021} and maps \citep[e.g.,][]{Friesen2010_3,Parise2011,Chacon-Tanarro+2019, Petrashkevich2024}. The highest deuterium fraction $R_D^{\rm N_2H^+}$ of 0.5--0.7 is observed toward starless cores in low-mass star-forming regions \citep{Crapsi2005,Pagani2007,Punanova2016}. Studies of deuterium fractionation toward protostars \citep{Emprechtinger2009,Pineda2010,Huang2015, Imai2018} show that fractionation continues after formation of a protostar that leads to the reverse process, a decrease of deuterium fraction. \citet{Emprechtinger2009} noted that protostars with a deuterium fraction $R_D^{\rm N_2H^+}$ = 0.15 are in a stage soon after the beginning of collapse, and protostars with a value of $\sim$0.03 are at the Class 0/I borderline. Deuterium fractionation was studied in relation to CO depletion, and while most of the results are in line with theoretical expectations \citep[e.g.,][]{Bacmann2003,Crapsi2005}, there are some outlying sources. \citet{Friesen2013} and \citet{Punanova2016} showed that an elevated deuterium fraction (0.12--0.43) was observed at a low CO depletion factor, $f_d$, (0.2--4.4) toward several dense cores in the Perseus and Ophiuchus MC, which contradicts the theory of deuterium fractionation \citep{Bacmann2003,Caselli2008,Kong2015}. The cores in Ophiuchus were further studied in \citet{Petrashkevich2024}.

One of the cores with high $R_D$ and low $f_d$ was a starless core in Barnard~5 (B5), an elongated structure located in the Perseus molecular cloud. B5 is located at a distance of 302$\pm$21 pc \citep{Zucker2018}. \citet{Pineda2010} showed a sharp transition from supersonic turbulence of the cloud to subsonic turbulence of the B5 filament in ammonia, a so-called transition to coherence. B5 hosts a Class~0--I protostar \citep[IRAS~03445+3242 or IRS1; see, e.g.,][]{Beichman1984,Yu1999} and three starless cores, two of them very close to IRS1; the cores may be forming a quadruple system \citep{Pineda2015}. According to the \ce{NH3} radial velocity and molecular hydrogen column density ($N$(\ce{H2})) analysis, B5 consists of two filaments, with average central densities of $\simeq1.5\times10^6$~cm$^{-3}$ \citep{Schmiedeke2021}, that accrete material \citep{Choudhury2024}. The maximum gas and dust temperature is $\simeq15$~K toward IRS1 \citep{Schmiedeke2021,Pezzuto2021}.  
In B5, deuterium fraction was studied only via pointing observations, toward B5 IRS1 \citep[in \ce{NH3}, \ce{H2CO}, \ce{N2H^+}, HNC;][respectively]{Hatchell_2003,Roberts2007,Friesen2013,Imai2018} and toward the dust peak of the starless core \citep{Friesen2013}, as well as CO depletion \citep{Friesen2013}. 


In this work, we present first maps of CO depletion and deuterium fraction in tracers of low- and high-density gas, \ce{HCO^+}, \ce{N2H^+}, and \ce{NH3} toward B5, to understand the distribution of deuterium fraction and its relation to CO depletion in this complex region. We analyse how protostar and CO depletion factor affect the deuterium fraction in the surrounding gas and explore the reason for the anomalously high deuterium fraction accompanied by low CO depletion in the B5 region, reported by \citet{Friesen2013}. We use the $^{13}$CO(1--0) emission \citep{Ridge2006} in the entire Perseus MC to estimate the total CO abundance in Perseus and obtain accurate values of $f_d$. 

Section \ref{sec:observations} presents the observations and the data taken from the literature. Section \ref{sec:observation_processing} presents the processing of the spectral maps. Section \ref{sec:results} describes the derived distributions of column densities, deuterium fractions, abundances and CO depletion factor. In Section~\ref{sec:discussion} we discuss the results. Section \ref{sec:conclusion} presents our conclusions.


\section{Observation Data}\label{sec:observations}
\subsection{IRAM 30~M Data}
\begin{figure}
\centering
\includegraphics[scale=0.35]{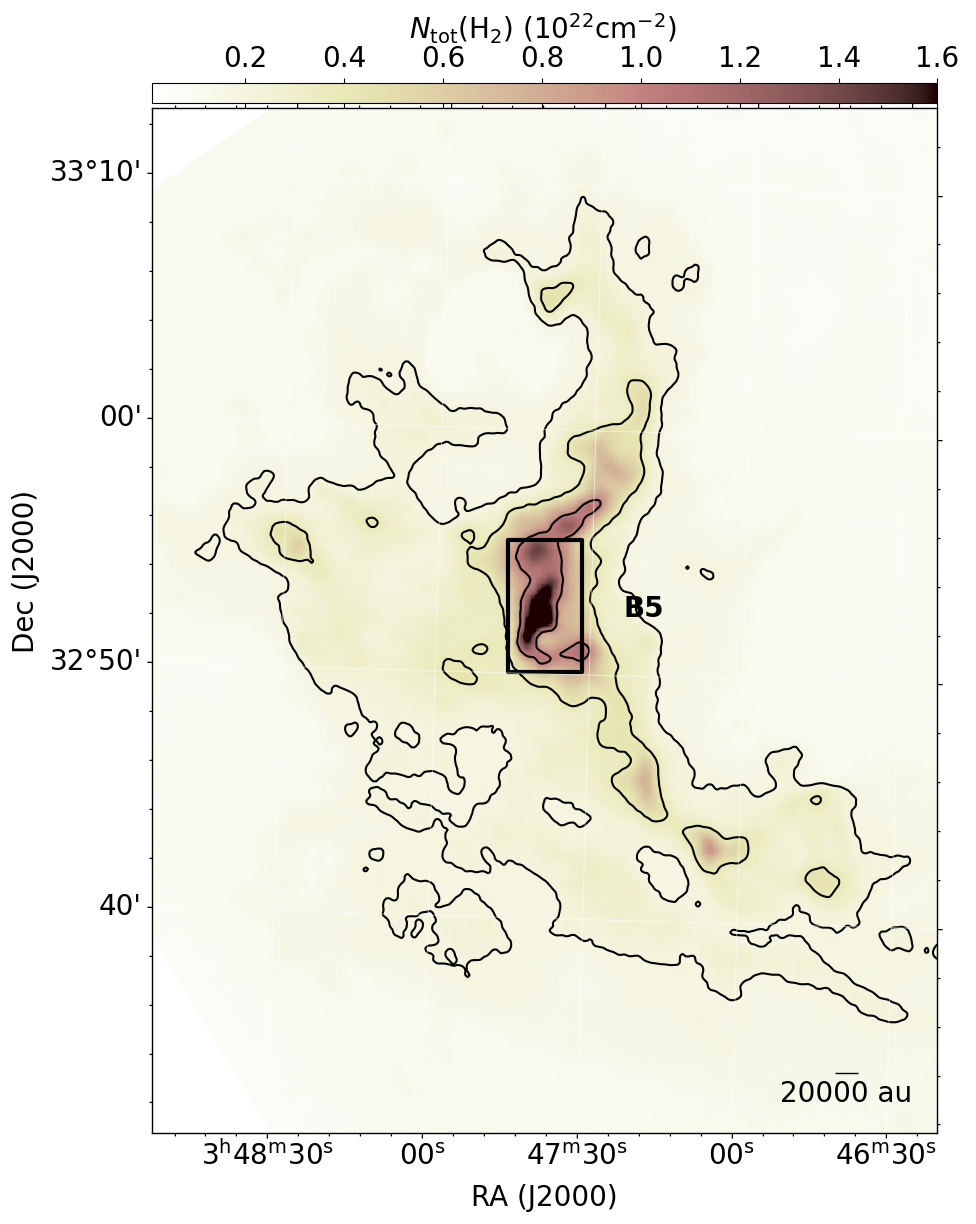}
\caption{Molecular hydrogen column density ($N_{\rm tot}$(H$_2$), color scale) toward B5 \citep[with the 36.3$^{\prime\prime}$ beam;][obtained through the Herschel Gould Belt Survey Archive]{Pezzuto2021}. The first contour starts at 0.1$\times$10$^{22}$~cm$^{-2}$ with a contour step of 0.5$\times$10$^{22}$~cm$^{-2}$. The black rectangle shows the area of our maps.}
\label{fig:Pers}
\end{figure}

Figure~\ref{fig:Pers} shows the B5 region with the area of our maps covering the brightest starless core (J2000 $\alpha$: 03$^h$ 47$^m$ 42.778$^s$, $\delta$: 32$^{\circ}$ 51$^{\prime}$ 30.31$^{\prime \prime}$) of B5 and the protostellar core with IRAS03445+3242 \citep[J2000 $\alpha$: 03$^h$ 47$^m$ 42.4$^s$, $\delta$: 32$^{\circ}$ 51$^{\prime}$ 44$^{\prime \prime}$,][]{Pineda2015}. The DCO$^+$(2--1), H$^{13}$CO$^+$(1--0), N$_2$D$^+$(1--0), and \ce{p-NH2D}(1,1) line observations were carried out with the IRAM-30m telescope on August 27--28 and November 23 2016 (project 031--16, PI: A.~Punanova). On-the-fly mapping was performed simultaneously in 2~and 3~mm bands, in position-switching mode, under acceptable weather conditions (PWV=7--10~mm). The telescope was equipped with the EMIR090 and EMIR150 receivers and the VESPA spectrograph, with a spectral resolution of 20 kHz. The telescope pointing and focus were checked every two and six hours, respectively, by observing QSO~0316+413, QSO~0430+052, or QSO~0234+285.

Observations of the N$_2$H$^+$(1--0) and C$^{18}$O(2--1) lines were obtained with IRAM 30~m on September 9--13 and November 5--6 and 16--17 2010 (project 025-10, PI: J.~Pineda). The observations were obtained with the EMIR090 and HERA1 Pixel1 receivers and the VESPA spectrograph, with a spectral resolution of 10~kHz and 39~kHz, respectively, in position-switching mode, under acceptable weather conditions (PWV=2--9~mm for the \ce{N2H^+} observations and PWV=0.5--3~mm for the C$^{18}$O observations). The telescope pointing and focus were checked every two and six hours, respectively, by observing QSO~0316+413, QSO~0420-014, QSO~0336+3218, W3OH, Uranus, or Mercury. The exact transition frequencies, spectral resolutions, beam sizes, antenna efficiencies, and system temperatures are listed in Table~\ref{tab:allline}. 

\subsection{Data from Literature}
We used maps of molecular hydrogen column density and dust temperature by \citet{Pezzuto2021}, available through the Herschel Gould Belt Survey Archive\footnote{Herschel Archive: \url{http://www.herschel.fr/cea/gouldbelt/en/index.php}}, with a beam size of 36.3$^{\prime\prime}$ and a pixel size of 3$^{\prime\prime}$. We used \ce{NH3} (1,1) and (2,2) maps from \citet{Pineda2010} obtained with the 100~m GBT, to estimate the gas temperature (the beam size of 31$^{\prime\prime}$, the pixel size of 15.5$^{\prime\prime}$, and the spectral resolution of 3.05~kHz). We use the $^{13}$CO(1--0) observations of the Perseus molecular cloud and L1688 star-forming region \citep[that was studied in][]{Petrashkevich2024} from the COMPLETE survey \citep{Ridge2006} obtained with the FCRAO antenna (the beam size of 46$^{\prime\prime}$, the pixel size of 23$^{\prime\prime}$, and the spectral resolution of 24~kHz) to estimate the total CO abundance in the clouds and pick the most suitable canonical CO abundance needed to calculate the CO depletion factor ($f_d$, see Sect.~\ref{sec:CO_depl} for details).

\begin{table*}
\caption{The list of the observed transitions, observation parameters, and the excitation temperatures ($T_{\rm ex}$) applied to calculate the column densities (see Sect.~\ref{sec:col_den}).}\label{tab:allline}
\begin{tabular}{lccccccccccc}
\hline\hline
Transition & Frequency &$T_{\rm sys}$ & HPBW$_{\rm obs}$  & $F_{\rm eff}$ & $B_{\rm eff}$ & rms, $T_{\rm mb}$ & $\Delta \varv_{\rm res}$ &$T_{\rm ex}$    \\ 
 & GHz & (K) & ($^{\prime\prime}$) &   & & (K)& (km s$^{-1}$) & (K) \\ \hline

DCO$^+$(2--1) & 144.0772804 & 185 & 17.1$^{\prime\prime}$  & 0.93 & 0.73& 0.07 & 0.041 &  5.2$\pm$0.6  \\
H$^{13}$CO$^+$(1--0)& ~86.7542884  &  177   &   28.4$^{\prime\prime}$ & 0.95 & 0.80 & 0.11&    0.067 &  6.0$\pm$0.8    \\
N$_2$D$^+$(1--0)& ~77.1096162   &    206   &  33.6$^{\prime\prime}$  & 0.95& 0.81& 0.09 & 0.076 & 4.5$\pm$0.7 \\
N$_2$H$^+$(1--0)  & ~93.1737637   &    --    &  26.4$^{\prime\prime}$ & 0.95 & 0.80 & 0.06 & 0.031 &   5.5$\pm$0.5  \\ 
$p$-NH$_2$D(1$_{11}^a$--1$_{01}^s$)  & ~110.153599   &    219    &  23.5$^{\prime\prime}$ & 0.95 &  0.79& 0.14  & 0.053 &  3.4$\pm$0.2  \\
NH$_3$(1$_{1}^a$,1$_{1}^s$)  & ~23.6958849   &    --    &  31.0$^{\prime\prime}$ & 0.95 & 0.80 & 0.02& 0.031 &   4.8$\pm$0.2  \\ 
NH$_3$(2$_{1}^a$,2$_{1}^s$)  & ~23.7240257   &    --    &  31.0$^{\prime\prime}$ & 0.95 & 0.80 & 0.02 & 0.031 &   4.8$\pm$0.2  \\ 
C$^{18}$O(2--1)& ~219.5603541   &    280   &  11.8$^{\prime\prime}$  & 0.95& 0.61 & 0.07& 0.053&  10 \\\hline
\end{tabular}
\begin{flushleft}
      \small
      \begin{flushleft}
      \item \textbf{Note:} $T_{\rm sys}$ is the system noise temperature. ${\rm HPBW}_{\rm obs}$ is the native half power beam width at the observed frequency. $F_{\rm eff}$ and $B_{\rm eff}$ are forward and beam efficiency of the antenna. The rms is given in the units of main beam temperature ($T_{\rm mb}$). $\Delta \varv_{\rm res}$ represents the spectral resolution in the units of velocity. For the data analysis, all cubes were convolved with the same 33.6$^{\prime\prime}$ beam. The transition frequencies and relative intensities of the hyperfine structure (HFS) components are taken from the following works: DCO$^+$(2--1) from \cite{Lattanzi2007_DCO21}, H$^{13}$CO$^+$(1--0) from \cite{Schmid-Burgk2004_H13CO}, N$_2$D$^+$(1--0) and N$_2$H$^+$(1--0) from \cite{Pagani2009_N2H_N2D}, $p$-NH$_2$D(1$_{11}^a$--1$_{01}^s$) from \citet{Daniel2016}, NH$_3$(1,1) and NH$_3$(2,2) from \citet{Harju2017}, and C$^{18}$O(2--1) from \cite{Klapper2001}. The frequencies and relative intensities of the HFS components are also available via CDMS \citep{CDMS}. NH$_3$ observations were taken from \citet{Pineda2010}.
      \end{flushleft}
\end{flushleft}
\end{table*}

\section{Data Reduction and Analysis}\label{sec:observation_processing}
\subsection{Spectral Cubes}
\begin{figure*}
\center{\includegraphics[width=1\linewidth]{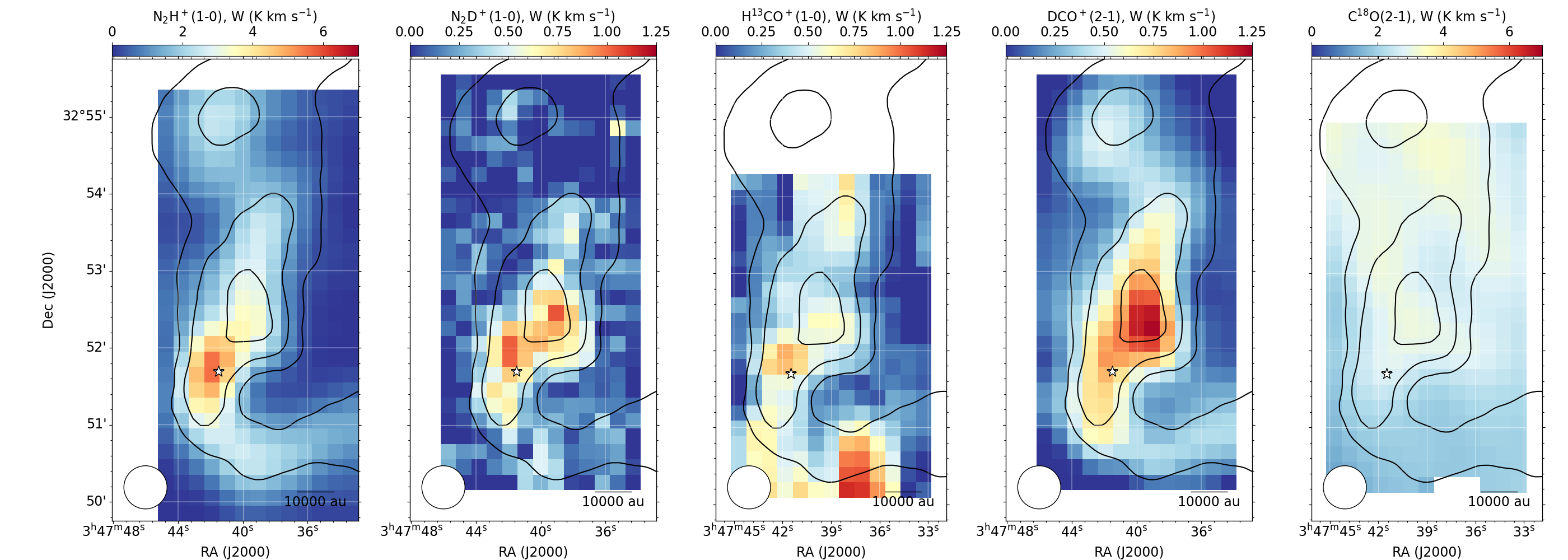}}
\caption{Integrated intensities ($W$) of the N$_2$H$^+$(1--0), N$_2$D$^+$(1--0), H$^{13}$CO$^+$(1--0), DCO$^+$(2--1), and C$^{18}$O(2--1) toward B5. The black contour shows the column density of molecular hydrogen. The first contour starts at 0.8$\times$10$^{22}$~cm$^{-2}$ with a contour step of 0.5$\times$10$^{22}$~cm$^{-2}$ \citep[with the 36.3$^{\prime\prime}$ beam;][obtained through the Herschel Gould Belt Survey Archive]{Pezzuto2021}. The star shows the position of the YSO \citep{Yu1999}. The beam size is shown in the bottom left corner of each map.}
\label{fig:W-maps}
\end{figure*}

We convolved our maps with the same angular resolution of 33.6$^{\prime\prime}$ that corresponds to the largest beam size in our dataset, \ce{N2D^+}(1--0), for consistency, and a pixel size of 12$^{\prime\prime}$ (consistent with the Nyquist criterion). Figure \ref{fig:W-maps} shows the integrated intensity ($W$) maps of the observed species. Up to the stage of the spectral cubes, the spectra were processed with \textsc{gildas} \footnote{\textsc{gildas}: \url{https://www.iram.fr/IRAMFR/GILDAS/}}. 

$N$(H$_2$) maps and the NH$_3$ (1,1) and (2,2) maps were regridded to match our 33.6$^{\prime\prime}$ beam and 12$^{\prime\prime}$ pixel. We also regridded the $N$(H$_2$) maps of the L1688 star-forming region and of the Perseus molecular cloud \citep{Ladjelate2020,Pezzuto2021} to match the beam and pixel sizes (46$^{\prime\prime}$ beam and 23$^{\prime\prime}$ pixel) of the $^{13}$CO maps from the COMPLETE survey. 

\subsection{Spectral Line Analysis}\label{sec:hfs-fit}
\subsubsection{HFS Analysis}
All lines that we use to estimate the deuterium fraction, \ce{N2H^+}(1--0), \ce{N2D^+}(1--0), \ce{H^{13}CO^+}(1--0), \ce{DCO^+}(2--1), \ce{NH3} (1,1) and (2,2) lines, and $p$-NH$_2$D(1$_{11}$--1$_{01}$) have hyperfine structures (HFS) that allow us to measure their optical depth $\tau$ (assuming the same excitation temperature, $T_{\rm ex}$, for all the hyperfines). The fitting of the \ce{NH3} lines employs an individual procedure and is described in Sect.~\ref{sec:ammonia}. To fit the other spectra, we used the package \textsc{pyspeckit}\footnote{\textsc{Pyspeckit}: \url{https://github.com/pyspeckit/pyspeckit}} \citep{Ginsburg2011,Ginsburg2022}. The package utilizes the frequencies and relative intensities of the HFS components to model the HFS under the assumption of local thermodynamic equilibrium, LTE, or common $T_{\rm ex}$ for all the hyperfines. The fitted parameters include the transition excitation temperature $T_{\rm ex}$, optical depth $\tau$, radial velocity $V_{\rm LSR}$, and velocity dispersion $\sigma$, related to the full width at half maximum (FWHM) as $\sigma = {\rm FWHM}/(2\sqrt{2\cdot\ln{2}})$. The package employs the gradient descent method and the radiation transfer equation to identify peaks in the spectrum and calculate the fit parameters. 

The map of $p$-NH$_2$D(1$_{11}$--1$_{01}$) had poor sensitivity, and after smoothing to 0.16~km~s$^{-1}$ and the convolution to 33.6$^{\prime\prime}$ beam, only a few pixels toward the dust peak of the starless core had the signal-to-noise ratio (S/N) $>3$. We thus analyse only the spectrum toward the dust peak with a free-parameter fit (the spectrum and the fit results are given in Fig.~\ref{fig:NH2D} in Appendix A).

We show and analyse only the pixels with relative uncertainty in parameter estimation less than 33\%. 
Among the four parameters in the fit, the optical depth and the excitation temperature have the highest uncertainty. To increase the number of pixels with low uncertainty, we explored the parameter space of $\tau$ and $T_{\rm ex}$ and their correlation between each other using the Markov chain Monte Carlo method, implemented in \textsc{pyspeckit} \citep{Ginsburg2011,Ginsburg2022} through \textsc{pymc} \citep{pymc2023}, with $10^5$ iterations, burnout of $10^3$, 250 iterations to find the parameter steps, and our fit parameters as priors,\footnote{See \url{https://pyspeckit.readthedocs.io/en/latest/example_pymc.html} for details.} as was done previously in \citet{Petrashkevich2024}. To explore the parameter space for all four lines, we started with the spectra with the highest S/N, toward the dust peak of the starless core and toward the YSO.
Figure \ref{fig:zone1} (in Appendix A) shows the parameter space exploration of optical depth and excitation temperature for the fitted lines. The optical depth has a larger range of values (about 9) than the excitation temperature (about 5~K) and affects the derived column densities, $N_{\rm tot}$, more than $T_{\rm ex}$. 
The column density is proportional to the optical depth while it increases with the $T_{\rm ex}$ by 21--27\% per 1~K for N$_2$H$^+$ and N$_2$D$^+$, by 12--19\% per 1~K for DCO$^+$ and H$^{13}$CO$^+$ at $T_{\rm ex}=4$--8~K \citep[see Sect.~3.2 in][]{Petrashkevich2024}.  With a free-parameter fit of the \ce{N2H^+}(1--0), \ce{N2D^+}(1--0), \ce{H^{13}CO^+}(1--0), and \ce{DCO^+}(2--1) lines, the $T_{\rm ex}$ toward the YSO and toward the starless core differed only by $\leq$1~K. Since the maps cover cold gas, we consider that the excitation conditions there are similar to those in the starless core, and since with our spatial resolution we do not resolve the gas heated by the YSO, we used the $T_{\rm ex}$ derived toward the starless core for the entire maps including the positions toward the YSO. We chose one excitation temperature value for each line and fixed it when fitting our spectra, to be able to estimate $\tau$ more accurately. The chosen excitation temperatures are given in Table~\ref{tab:allline}.

We removed pixels with a relative uncertainty of $>33$\% from the respective parameter maps. Figure \ref{fig:spec} (in Appendix A) shows the fitted spectra for all lines toward the cold dense core and the YSO. We subtracted the spectrum model from the observed spectrum to estimate the accuracy of each fit. The difference between the noise rms and the rms of the subtraction was within 8\%. 

\subsection{\ce{NH3}, \ce{C^{18}O}, $^{13}$CO Data}\label{sec:ammonia}

To calculate the gas temperature, we used the observational maps of para-NH$_3$(1,1) and (2,2) from \citet{Pineda2010}. Under molecular cloud conditions ($n\sim10^3-10^6$~cm$^{-3}$, $T\sim10-100$~K), the excitation of the lower metastable inversion transitions of \ce{NH3} is dominated by collisions \citep[e.g.,][]{Ho1983}, which makes the \ce{NH3} lines an excellent temperature tracer. We used the model of ammonia spectra that fits both transitions, implemented in \textsc{pyspeckit} by \cite{Ginsburg2011} building up on \citet{Rosolowsky_2008}, which takes into account the ortho-to-para ($o/p$) ratio for NH$_3$ as a parameter. We used $o/p$~=~1:1, following \citet{Harju2017}. The method estimates the following spectral parameters: optical depth, radial velocity, velocity dispersion, excitation temperature, and the sum column density of ortho- and para- \ce{NH3}. For the fit, we used the spectra with the S/N $>5$. The difference between the fit and the spectrum was 7--10\%. The excitation temperature is then adapted as the gas kinetic temperature (see left panel of Fig.~\ref{fig:T_g} in Appendix A). 

We applied a Gaussian fit to the \ce{C^{18}O}(2--1) line to estimate the radial velocity and velocity dispersion. The rightmost panels of Fig.~\ref{fig:tau} (in Appendix A) show the radial velocity and velocity dispersion maps obtained from the fitting. All pixels have the uncertainty of $V_{\rm LSR}$ in the range of 0.02--0.04~km~s$^{-1}$ and that of $\sigma$ in the range of 0.02--0.03~km~s$^{-1}$ which means the relative uncertainty $<7$\%. Lastly, to obtain the integrated intensity of $^{13}$CO(1--0), we used only the pixels with the S/N $>3$. 

\section{Results}\label{sec:results}

\subsection{Fit Parameters}\label{sec:spparam}

The integrated intensity maps (see Fig.~\ref{fig:W-maps}) and the line profiles (see Fig.~\ref{fig:spec} in Appendix A) showed that the line intensity of the hydrogen-bearing species is higher toward the protostellar core and the line intensity of the deuterated species is higher toward the starless core. 

Figure \ref{fig:tau} (in Appendix A) shows the HFS fit parameters: the optical depth $\tau$, the radial velocity $V_{\rm LSR}$ and the velocity dispersion $\sigma$. The excitation temperatures used for the fit are shown in Table~\ref{tab:allline}. The number of pixels in the parameter maps is limited by their relative uncertainties ($\leq30$ \%). The uncertainties of the radial velocities were 
0.01--0.02~km~s$^{-1}$. The radial velocity varied from 9.95 km~s$^{-1}$ to 10.41 km~s$^{-1}$, with the maximum value reached to the south-west of the starless core. The values of radial velocity and velocity dispersion in \ce{N_2H^+} and \ce{N_2D^+}, and the velocity pattern agree with that of N$_2$H$^+$ and NH$_3$ in \citet{Pineda2021,Schmiedeke2021}.

The velocity dispersion shows the values 0.07--0.32~km~s$^{-1}$ across the map. The median difference between the dispersions of \ce{N2H^+}(1--0) and \ce{N2D^+}(1--0) is 0. The \ce{DCO^+}(2--1) and \ce{H^{13}CO^+}(1--0) lines are wider than \ce{N2H^+}(1--0) on median by 0.02~km~s$^{-1}$ and 0.05~km~s$^{-1}$, respectively. The largest values of 0.18--0.32~km~s$^{-1}$ are observed toward the protostellar core in \ce{H^{13}CO^+}(1--0). This indicates 
that \ce{H^{13}CO^+} traces a larger fraction of the gas on the line of sight. In the other lines, the largest dispersion is observed toward the south-west of the starless core. The \ce{H^{13}CO^+}(1--0) line also shows hints of a second velocity component toward the protostellar core and to the south-east from it. It probably traces the interface of two colliding cores \citep[see][for the details]{Schmiedeke2021}. 

\begin{figure}
\centering
\includegraphics[scale=0.5]{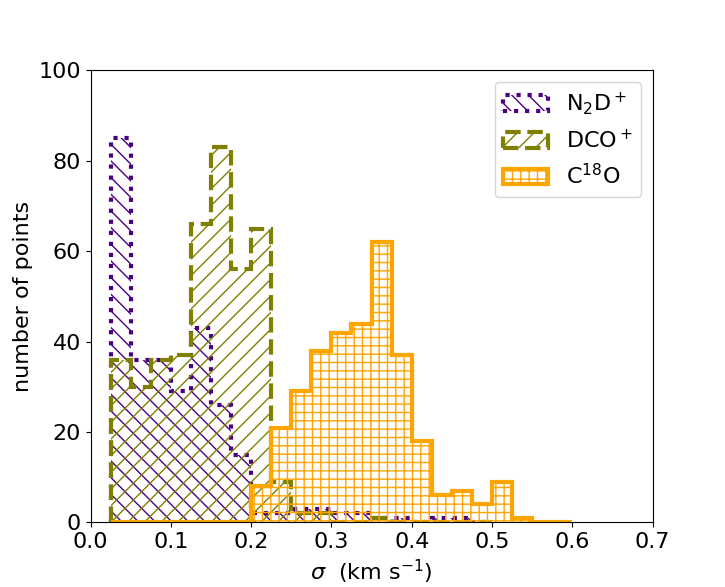}
\caption{Distribution of the velocity dispersion $\sigma$ of \ce{N2D^+}(1--0), \ce{DCO^+}(2--1), and C$^{18}$O(2--1). The step is 0.025~km~s$^{-1}$.} 
\label{fig:sigma}
\end{figure}

The rightmost panels of Fig.~\ref{fig:tau} (in Appendix A) show the radial velocity and the velocity dispersion of the \ce{C^{18}O}(2--1) line. The \ce{C^{18}O} radial velocity differs from the radial velocity of the other species because \ce{C^{18}O} traces mainly less dense gas of the surrounding cloud, as illustrated by the factor of $\simeq$2 larger velocity dispersions compared to that of the deuterated species tracing the dense gas (see Fig.~\ref{fig:sigma}). The distribution of the \ce{C^{18}O} velocity dispersion differs from those of the other species, with the largest values of 0.53~km~s$^{-1}$ observed toward another dust peak to the north from both cores, where the velocity dispersion of \ce{N2H^+}(1--0) and \ce{DCO^+}(2--1) is 0.10--0.17~km~s$^{-1}$. The radial velocity pattern of \ce{C^{18}O}(2--1) differs from that of the other species, with $V_{\rm LSR}$ increasing from the north-west to south-east of the map, and the values are overall lower than, e.g., $V_{\rm LSR}$ of \ce{N2H^+}(1--0) by -0.2--0.4~km~s$^{-1}$ (on median by 0.08~km~s$^{-1}$). The differences in $\sigma$ and $V_{\rm LSR}$ between \ce{C^{18}O} and the dense gas tracers should be first of all accounted to the \ce{C^{18}O}(2--1) line profiles that show the presence of unresolved or partly resolved second velocity component that probably traces the gas of the surrounding cloud. Since kinematics is out of scope of this paper, we do not analyse them separately. The hints of the second velocity component are mostly present toward the north from the starless core; they contribute to measured column density and may lead to slight underestimate of CO depletion factor (see Sect.~\ref{sec:col_den} and~\ref{sec:CO_depl} for details).

\subsection{Column Density and Abundance}\label{sec:col_den}
For the lines with the HFS, we used the fit parameters to calculate the column density assuming that the emission is consistent with LTE \citep[see, e.g.,][]{Shirley2015}. We calculate the column density as it was done in \citet{Caselli2002}: 
\begin{equation}
N_{\rm tot}=\frac{8\pi^{3/2}\sigma}{\lambda^3A_{\rm ul}} \frac{g_l}{g_u} \frac{\tau}{1-\exp(-h\nu/kT_{\rm ex})} \frac{Q_{\rm rot}}{g_l\exp(-E_l/kT_{\rm ex})},
\end{equation}
where $\sigma$ is the velocity dispersion, $\lambda$ is the transition wavelength, $A_{\rm ul}$ is the Einstein coefficient, $g_l$ and $g_u$ are the statistical weights of the lower and upper energy levels, $\tau$ is the optical depth, $h$~is the Planck constant, $\nu$~is the transition frequency, $k$ is the Boltzmann constant, $T_{\rm ex}$~is the excitation temperature, $Q_{\rm rot}$ is the partition function, $E_l$ is the energy of the lower level. The partition function of a linear molecule is calculated as follows:
\begin{equation}
Q_{\rm rot}=\sum^\infty_{J=0}(2J+1)\exp(E_J/kT_{\rm ex}),
\end{equation}
where $J$ is the rotational quantum number, and $E_J$ is the rotational transition energy. 
The constants for the studied transitions are the same as those used in \citet{Petrashkevich2024}, see Table~3 there.
We applied the $^{12}$C/$^{13}$C=77 isotopic ratio for the local ISM \citep{Wilson1994_HCO} to calculate the HCO$^+$ column density from our $N$(H$^{13}$CO$^+$) data. 

\begin{figure*}
\centering
\includegraphics[scale=0.25]{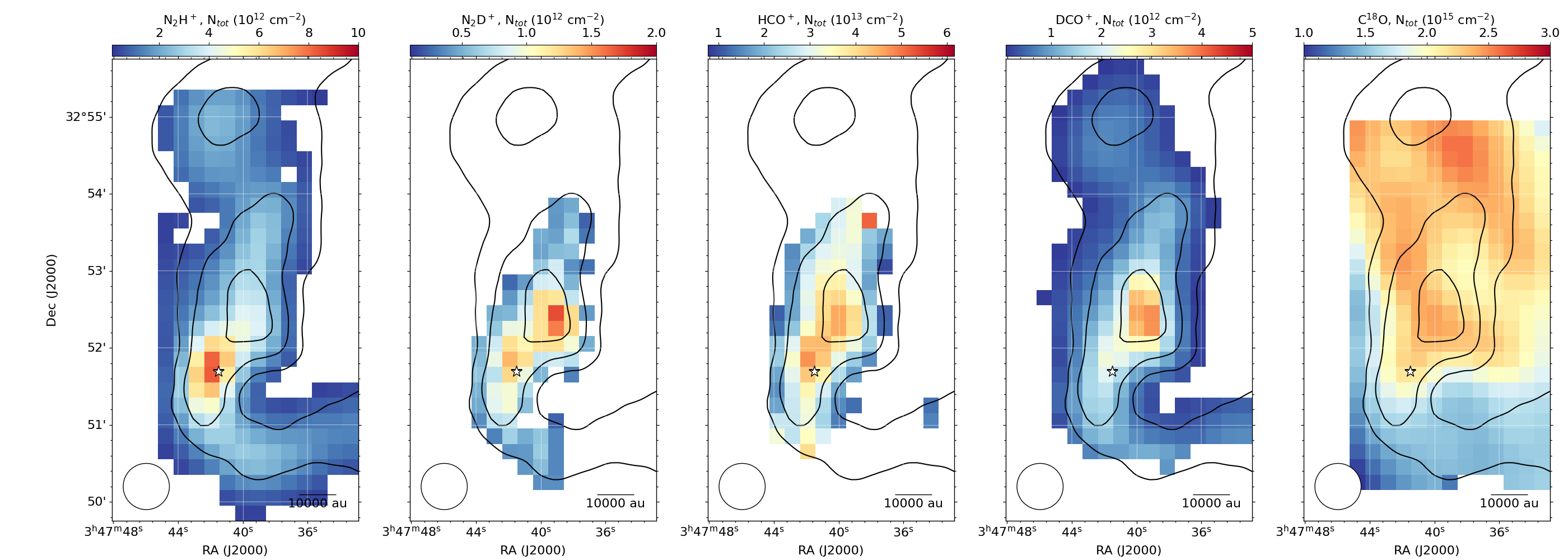}
\caption{The column density maps of N$_2$H$^+$, N$_2$D$^+$, DCO$^+$,HCO$^+$, and C$^{18}$O toward B5. The black contour shows the column density of molecular hydrogen (the first contour starts at 0.8$\times$10$^{22}$~cm$^{-2}$ with a contour step of 0.5$\times$10$^{22}$~cm$^{-2}$). The star shows the position of the YSO \citep{Yu1999}. The beam size is shown in the bottom left corner of each map.}
\label{fig:N}
\end{figure*}

Figure \ref{fig:N} shows the column density maps calculated from the lines with HFS.  We obtained the following estimates of the column densities: 0.4--8.3$\times$10$^{12}$~cm$^{-2}$ of N$_2$H$^+$; 0.5--1.9$\times$10$^{12}$~cm$^{-2}$ of N$_2$D$^+$; 1.0--4.5$\times$10$^{13}$~cm$^{-2}$ of HCO$^+$; 0.5--3.7$\times$10$^{12}$~cm$^{-2}$ of DCO$^+$. The column density of p-\ce{NH2D} toward the starless core is $(2.6\pm1.4)\times10^{13}$~cm$^{-2}$, which gives $N$(\ce{NH2D})=$(1.04\pm0.56)\times10^{14}$~cm$^{-2}$, assuming o/p ratio of \ce{NH2D}=3 following \citet{Harju2017}. 
The column density peaks of N$_2$H$^+$ and HCO$^+$ appear toward the protostellar core, and the column density peaks of N$_2$D$^+$ and DCO$^+$ appear toward the starless core. The column densities of N$_2$D$^+$ and DCO$^+$ toward the starless core are larger than those toward the protostellar core by factors of 1.3 and 1.8, respectively. HCO$^+$ and N$_2$D$^+$ show secondary column density peaks toward the starless (HCO$^+$) and the protostellar (N$_2$D$^+$) cores. 

\cite{Friesen2013} measured the column densities of N$_2$D$^+$ and N$_2$H$^+$ toward three positions in B5 (their 3 to the north from the YSO, 4 toward the YSO, and 6 toward the starless core). Our column density of N$_2$D$^+$ toward positions 3 and 6 is similar to theirs within the uncertainty; toward position 4 our $N$(\ce{N2D^+}) is higher by a factor of three. The latter may be because \citet{Friesen2013} detected only one, higher, transition \ce{N2D^+}(3--2) and considered it optically thin. However, our column density of N$_2$H$^+$ in those positions is 2.7--8.3$\times$10$^{12}$~cm$^{-2}$, that is 70\% lower than that of \cite{Friesen2013}. \citet{Emprechtinger2009} also obtained a factor of 2 larger $N$(\ce{N2H^+}) toward B5 IRS1 and lower $N$(\ce{N2D^+}). The reason is our convolution of the N$_2$H$^+$ data with a larger beam and the fact that they used a higher transition of \ce{N2D^+}(2--1) and considered it optically thin. 

The C$^{18}$O(2--1) line has no HFS. \citet{Pineda2008} in their study of CO in Perseus (including B5) concluded that C$^{18}$O(1--0) is optically thin there; \citet{Friesen2013} considered the C$^{18}$O(2--1) lines as optically thin too. Thus we calculated the column density of C$^{18}$O(2--1) as was done in \citet{Caselli2002} for optically thin transitions:
\begin{multline}
N_{\rm tot} =\frac{8W\pi g_l}{\lambda^3A_{\rm ul}g_u} \frac{1}{J_\nu(T_{\rm ex})-J_\nu(T_{\rm bg})}\times\\  \times \frac{1}{1-\exp(-hv/kT_{\rm ex})}\frac{Q_{\rm rot}}{g_l\exp(-E/kT_{\rm ex})},
\end{multline}
where  $W$ is the integrated intensity, $J_\nu(T)$ is the equivalent Rayleigh-Jeans temperature, $T_{\rm bg}$ is the background temperature of 2.7~K, $T_{\rm ex}$ is excitation temperature that we assume to be 10~K.
We estimated the C$^{18}$O column density and obtained 1.1--2.6$\times$10$^{15}$~cm$^{-2}$ $\pm2.8\times10^{13}$~cm$^{-2}$ over the map. In many positions, the C$^{18}$O(2--1) spectra show two unresolved or partly resolved velocity components separated by 0.5--1.0~km~s$^{-1}$ which is consistent with the presence of two filaments. We obtain the total column density integrating all components since we use $N$(H$_2$) estimated via the dust observations that is an integral over all velocities, to estimate CO abundance and CO depletion factor (see Sect.~\ref{sec:CO_depl}). Our estimate of the column density agrees within the uncertainty with the previous studies \citep{Pineda2008, Friesen2013}. We used the fractional abundance [$^{16}$O]/[$^{18}$O]=560 \citep{Wilson1994_HCO} to convert the column density of C$^{18}$O to that of CO considering that \ce{C^{18}O}/CO fractionation is insignificant on the scale of B5. The rightmost panel of Fig.~\ref{fig:N} shows the column density of C$^{18}$O.

To estimate the $^{13}$CO column density toward the Perseus molecular cloud and L1688, we used the \ce{^{13}CO}(1--0) data of the COMPLETE survey \citep{Ridge2006}. The \ce{^{13}CO}(1--0) line has HFS; however, the spectra often have blended multiple velocity components, and we obtained poor HFS fits. \citet{Pineda2008} estimated average optical depth of the \ce{^{13}CO}(1--0) line for each of the studied regions to be 0.3--0.4. We expect that CO is undepleted or least depleted in the most diffuse regions, and the \ce{^{13}CO}(1--0) lines are optically thin there. Thus, we used the same method as for \ce{C^{18}O}(2--1), assuming the lines were optically thin. Since the spectra have several velocity components \citep{Pineda2008}, we took the integrated intensity over all components to estimate all $^{13}$CO in the gas. For the regions in Perseus, we use $T_{\rm ex}$=9--12~K obtained by \citet{Pineda2008} separately for each region; for L1688, we set $T_{\rm ex}$=10~K. To convert the \ce{^{13}CO} column density to that of CO, we used $^{12}$C/$^{13}$C fractional abundance of 77 \citep{Wilson1994_HCO}. The results are discussed in detail in Sect.~\ref{sec:CO_depl}.


\subsubsection{Abundances}
Figure~\ref{fig:X} (in Appendix A) shows the abundance maps of \ce{N2H^+}, \ce{N2D^+}, \ce{HCO^+}, \ce{DCO^+}, and \ce{C^{18}O}. We estimated the abundances as follows: 
\begin{equation}
X=N_{\rm tot}/N_{\rm tot}(\text{H}_2),
\end{equation}
where $N_{\rm tot}$ is the column density of the species and $N_{\rm tot}$(H$_2$) is the column density of molecular hydrogen  from the Herschel Gould Belt Survey \citep[][the peak $N_{\rm tot}$(H$_2$) in B5 is 2.46$\times10^{22}$~cm$^{-2}$]{Pezzuto2021}. The abundance maps of \ce{N_2H^+} and \ce{DCO^+} are similar to the column density maps, with abundance peaks observed at the same locations. The \ce{HCO^+} abundance distribution differs from that of the column density, with no secondary peak toward the starless core. 
The abundances of \ce{N_2D^+} toward the protostellar and starless cores are similar. The median abundances over the maps are $X$(\ce{N2H^+})=1.47$\times$10$^{-10}$; $X$(\ce{N2D^+})=4.42$\times$10$^{-11}$; $X$(\ce{HCO^+})=1.85$\times$10$^{-9}$; $X$(\ce{DCO^+})=5.49$\times$10$^{-11}$. 
The \ce{C^{18}O} abundance decreases by a factor of 1.5-2 when the column density of molecular hydrogen increases up to 1.3$\times$10$^{22}$~cm$^{-2}$ (see black contour in the rightmost panel of Fig.~\ref{fig:X} in Appendix A). \citet{Pineda2008} estimated the average abundance of CO in B5 to be 2.1$\times$10$^{-4}$ which yields average $X$(\ce{C^{18}O})=3.75$\times$10$^{-7}$ when $^{16}$O/$^{18}$O=560 \citep{Wilson1994_HCO} is applied. Our estimate gives the average [$X$(\ce{C^{18}O})]=1.26$\times$10$^{-7}$ (the median value is the same) over our map, which has smaller area than the map of \citet{Pineda2008} and does not cover the area of undepleted CO. 

\begin{table*}
\caption{Column densities and deuterium fractions toward the protostellar and the starless core.}\label{tab:column} 
\begin{center}
\begin{tabular}{lccccccccc}
\hline\hline
Core& $N_{\rm tot}$(DCO$^+$)  & $N_{\rm tot}$(H$^{13}$CO$^+$) & $R_D^{\rm HCO^+}$ & $N_{\rm tot}$(NH$_2$D) & $N_{\rm tot}$(NH$_3$) &$R_D^{\rm NH_3}$ & $N_{\rm tot}$(N$_2$D$^+$) &$N_{\rm tot}$(N$_2$H$^+$)& $R_D^{\rm N_2H^+}$ \\
 & 10$^{12}$ (cm$^{-2}$) & 10$^{12}$ (cm$^{-2}$) & & 10$^{14}$ (cm$^{-2}$) & 10$^{14}$ (cm$^{-2}$) & & 10$^{12}$ (cm$^{-2}$) & 10$^{12}$ (cm$^{-2}$) &  \\ \hline
Protostellar & 2.11$\pm$0.35 & 0.55$\pm$0.08 & 0.049$\pm$0.011 & -- & 3.78$\pm$0.15 & -- & 1.27$\pm$0.19 & 8.31$\pm$1.13 & 0.15$\pm$0.03 \\ 
Starless & 3.75$\pm$0.61 & 0.53$\pm$0.09 & 0.094$\pm$0.024 & 1.04$\pm$0.56 & 2.92$\pm$0.06 & 0.36$\pm$0.20 & 1.71$\pm$0.27 & 3.95$\pm$0.59 & 0.43$\pm$0.10 \\ 
\hline
\end{tabular} 
\end{center}
\end{table*}

\subsection{CO Depletion} \label{sec:CO_depl}

The CO depletion factor, $f_d$, shows how much CO is adsorbed onto the dust grains relative to all CO present in the cloud. 
We calculated the CO depletion factor as follows:
\begin{equation}
f_d({\rm CO}) =\frac{X({\rm CO})_{\rm can}}{X({\rm CO})_{\rm obs}},
\end{equation}
where $X$(CO)$_{\rm can}$ is the canonical CO abundance and $X$(CO)$_{\rm obs}$=$N_{\rm tot}$(CO)/$N_{\rm tot}$(H$_2$). The canonical CO abundance (X(CO)$_{\rm can}$) with respect to H$_2$ in the interstellar medium was estimated in a number of studies: 0.39$\times$10$^{-4}$ \citep{Tafalla2004}, 0.95$\times$10$^{-4}$ \citep{Frerking1982}, 1.7$\times$10$^{-4}$ \citep{Wannier1980}, and 2.69$\times$10$^{-4}$ \citep{Lacy1994}. 


\begin{table*}
\caption{Minimum, median and maximum (min, med, max) $^{13}$CO depletion factor obtained with different canonical abundances toward Perseus molecular cloud (PMC) and L1688. $N$ shows the number and percentage of points with $f_d<1$.} \label{tab:depl}
\begin{center}
\begin{tabular}{ccccccccccccl}
\hline\hline
 $X_{\rm can}$(CO) & $X_{\rm can}$($^{13}$CO) & \multicolumn{3}{c}{$f_d$, PMC} & \multicolumn{3}{c}{$f_d$, L1688} & \multicolumn{2}{c}{$N(f_d<1)$, PMC} & \multicolumn{2}{c}{$N(f_d<1)$, L1688} & Citation \\ 
\cmidrule(lr){3-5} \cmidrule(lr){6-8} \cmidrule(lr){9-10} \cmidrule(lr){11-12}
   &   & min & med & max & min & med & max &  &  &  &  &  \\ \hline
2.69$\times$10$^{-4}$   & 3.5$\times$10$^{-6}$  & 0.94 & 2.52  & 47.03 & 0.43 & 3.14  & 50.31 & 6 & 0.1\% & 3306 & 3.5\% & \citet{Lacy1994} \\ 
 1.68$\times$10$^{-4}$   & 2.2$\times$10$^{-6}$  & 0.58 & 1.57 & 29.39 & 0.27 & 1.96 & 31.44 & 6953 & 10\% & 18440 & 20\% & \citet{Wannier1980} \\ 
 0.95$\times$10$^{-4}$   & 1.2$\times$10$^{-6}$  & 0.33 & 0.89 & 16.66 & 0.15 & 1.12 & 17.82 & 42150 & 61\% & 42150 & 45\% & \citet{Frerking1982} \\ 
 0.39$\times$10$^{-4}$   & 5.1$\times$10$^{-7}$  & 0.13 & 0.36 & 6.85 & 0.06 & 0.45 & 7.35 & 66248 & 96\% & 82077 & 88\% & \citet{Tafalla2004} \\ \hline
\end{tabular} 
\end{center}
\end{table*}

\begin{figure*}
\centering
\includegraphics[scale=0.39]{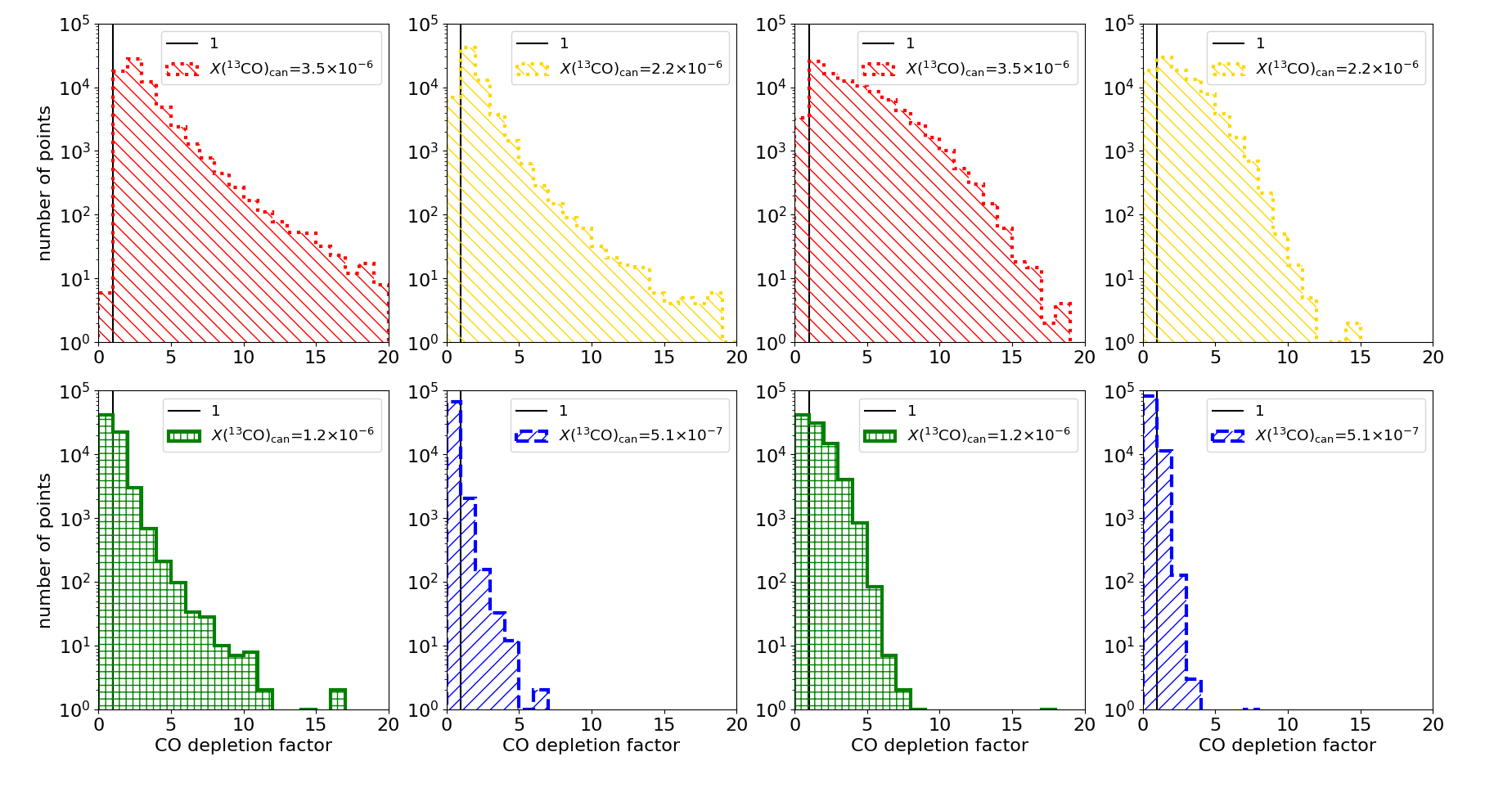}
\caption{$^{13}$CO depletion ($f_d$), based on the \ce{^{13}CO}(1--0) data \citep[the COMPLETE survey,][]{Ridge2006} toward the entire Perseus molecular cloud and L1688 region, obtained with different CO canonical abundances. The canonical CO abundances were taken from \citet[][$X$(\ce{CO})= 2.69$\times$10$^{-4}$, red]{Lacy1994}, \citet[][$X$(\ce{C^{18}O})=3.0$\times10^{-7}$, yellow]{Wannier1980}, \citet[][$X$(\ce{C^{18}O})=1.7$\times10^{-7}$, green]{Frerking1982}, \citet[][$X$(\ce{C^{18}O})=0.7$\times10^{-7}$, blue]{Tafalla2004}. The black line shows $f_d$ = 1 at which all CO is in the gas phase.} 
\label{fig:13CO}
\end{figure*}

To estimate the total CO abundance and choose the most suitable canonical CO abundance, we used $^{13}$CO(1--0) observations from the COMPLETE survey \citep{Ridge2006}. 
Figure \ref{fig:13CO} shows the distributions of the CO depletion factors obtained for the entire Perseus molecular cloud (left panels) and the L1688 star-forming region in the Ophiuchus molecular cloud (right panels) using different canonical CO abundances. The black line shows the CO depletion factor $f_d$ = 1. By definition, the points with $f_d<1$ are unphysical. Our choice of the canonical CO abundance is based on the number of such points. We chose the canonical CO abundance of $X$(CO)$_{\rm can}$ = 2.69$\times$10$^{-4}$ with respect to H$_2$ from \citet{Lacy1994} \citep[or 3.5$\times$10$^{-6}$ of $^{13}$CO, or 4.8$\times$10$^{-7}$ of C$^{18}$O if converted using the isotope ratios from][]{Wilson1994_HCO}, as it yields the least number of unphysical values compared to the others, both in Perseus and in L1688. The summary of $f_d$ distribution obtained with different $X_{\rm can}$(CO) is given in Table~\ref{tab:depl}. Since the $X_{\rm can}$(CO) were obtained for CO and C$^{18}$O, the applied isotope ratios may affect our result. While $^{16}$O/$^{18}$O=560 \citep{Wilson1994_HCO} is well established for the local ISM, there is more discrepancy in the reported $^{12}$C/$^{13}$C \citep[see, e.g.,][]{Milam2005}. With the increasing $^{12}$C/$^{13}$C ratio, the histograms in Fig.~\ref{fig:13CO} would move to the left, confirming the chosen $X_{\rm can}$(CO). With the decreasing $^{12}$C/$^{13}$C ratio, the histograms would move to the right, however, down to $^{12}$C/$^{13}$C=60 the chosen $X_{\rm can}$(CO) = 2.69$\times$10$^{-4}$ is still valid. In addition to galactocentric trends, the $^{12}$C/$^{13}$C ratios of various species depend on time and physical conditions. For CO, models show that the ratio starts at $\simeq10$ and increases with time evolution, stabilizes at $\sim10^5$~yr and tends to follow the elemental $^{12}$C/$^{13}$C ratio \citep{Colzi2020,Sipila2023}. The same canonical value was found to be the most suitable for the Taurus molecular cloud \citep{Punanova2022} based on C$^{18}$O data. This canonical CO abundance was used in other studies of star-forming regions \citep[e.g.,][]{Lee2003,Hernandez2011,SanJose-Garcia2013,Petrashkevich2024}. 

\begin{figure}
\centering
\includegraphics[scale=0.5]{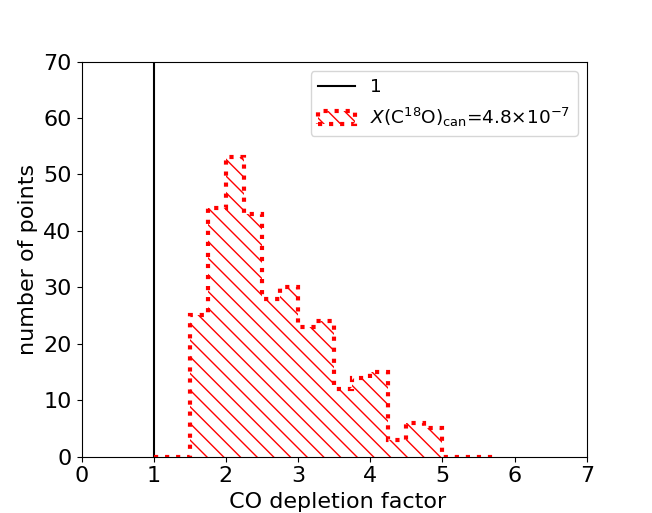}
\caption{CO depletion ($f_d$) based on our \ce{C^{18}O}(2--1) data toward B5. The canonical C$^{18}$O abundance (4.8$\times$10$^{-7}$) was taken from \citet{Lacy1994}. The black line shows $f_d$ = 1 at which all CO is in the gas phase.} 
\label{fig:C18O}
\end{figure}

To estimate how CO depletion affects the deuterium fraction in B5, we used the  C$^{18}$O(2--1) observations. We applied the chosen $X$(CO)$_{\rm can}=2.7\times10^{-4}$ \citep{Lacy1994} and the $^{16}$O/$^{18}$O=560 \citep{Wilson1994_HCO} to estimate $f_d$ in B5. 
Figure \ref{fig:C18O} shows the distribution of the CO depletion factor toward B5 and the center right panel of Fig.~\ref{fig:RD_fd} shows the map of CO depletion. The CO depletion factor varies between 1.2 and 5.0, with 
the maximum $f_d$ observed toward the starless core. CO depletion factor increases with $N$(\ce{H2}). Our estimate is higher than that obtained in \citet{Friesen2013} by a factor of 3--8, however, this is due to the choice of a different canonical CO abundance \citep[they used 1.7$\times$10$^{-7}$;][]{Frerking1982} and our smoothing with a larger beam, since we obtained similar column densities (see Sect.~\ref{sec:discus_depletion}).

We compared our estimate to the CO depletion factors obtained with the C$^{18}$O (1--0) and (2--1) lines in other regions of low-mass star formation. \citet{Crapsi2005} presented $f_d$ toward 13 cores in Taurus, Serpens, Aquila rift, and Ophiuchus that varied in the range of $f_d$=3.4--15.5. \citet{Crapsi2005} report higher $f_d$=11--15.5 for the cores with $N$(\ce{H2}) of (6--13.5)$\times 10^{22}$~cm$^{-2}$. For B5, we obtained $f_d$=5.0, smaller than the cores of \citet{Crapsi2005} in general, and closer to those of the Taurus cores ($f_d$=7.0--9.5) with similar $N$(\ce{H2}) of (1--4)$\times 10^{22}$~cm$^{-2}$. The small variation could be explained by the beam size: the 33.6$^{\prime\prime}$ beam in this work and 22$^{\prime\prime}$ and 11$^{\prime\prime}$ beams of C$^{18}$O (1--0) and (2--1), respectively, in \citet{Crapsi2005}. In addition to that, Perseus is at least factor of 2 more distant than Taurus and the linear size probed is larger in Perseus than in Taurus which also can affect the observed CO depletion factor.

\subsection{Deuterium Fraction}\label{sec:rd}

\begin{figure*}
\centering
\includegraphics[scale=0.32]{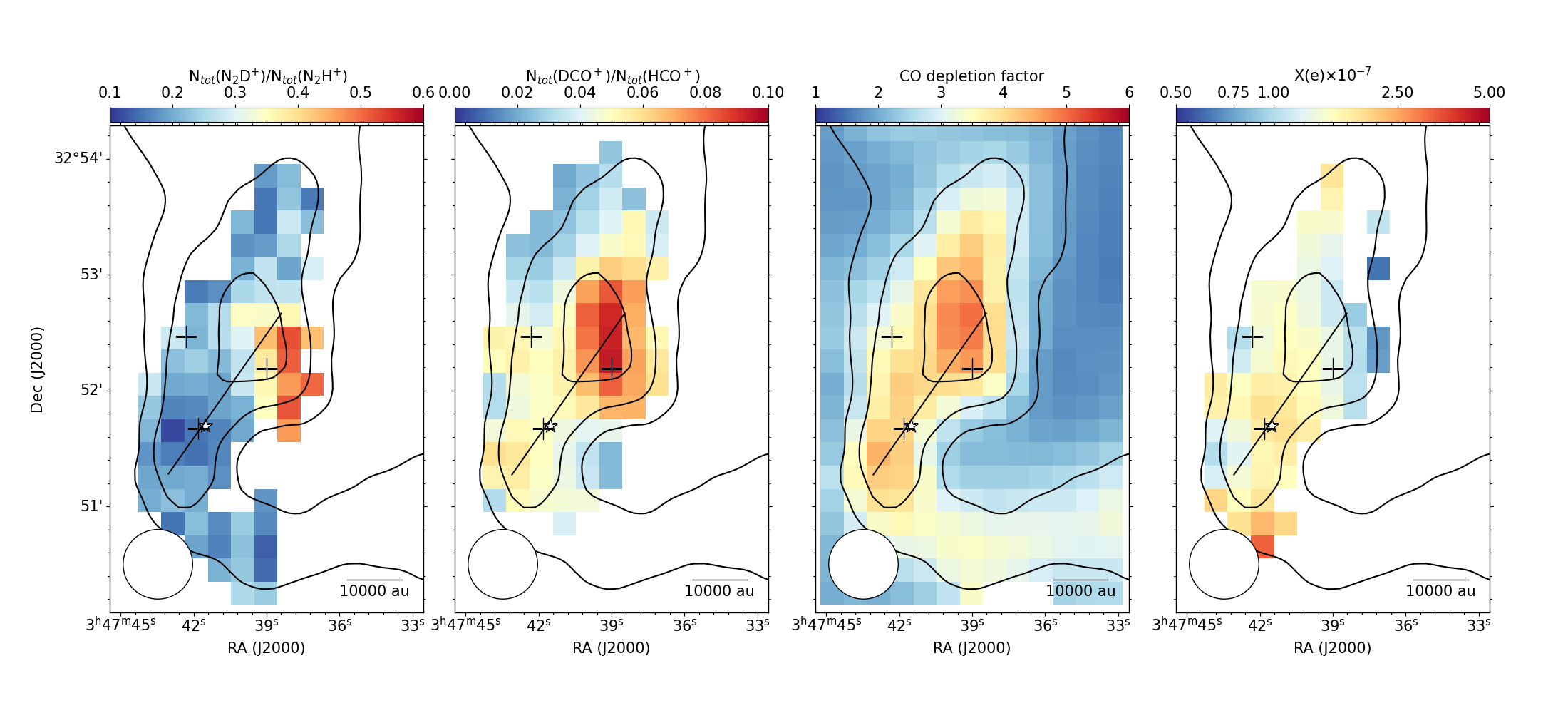}
\caption{The deuterium fraction maps of \ce{N2H^+} ({\it far left}) and \ce{HCO^+} ({\it center left}) toward B5. CO depletion toward B5 ({\it center right}). Ionization degree estimated from the sum of the ion abundances (\ce{N2H^+}, \ce{N2D^+}, \ce{HCO^+}, and \ce{DCO^+}, {\it far right}). The molecular hydrogen column density is shown by black contours (the first contour starts at 0.8$\times$10$^{22}$~cm$^{-2}$ with a contour step of 0.5$\times$10$^{22}$~cm$^{-2}$). The crosses show the positions of the single pointing observations in \citet{Friesen2013}. The star shows the position of the YSO \citep{Yu1999}. The beam size is shown in the bottom left corner of each map. CO depletion was calculated with 4.8$\times$10$^{-7}$ from \citet{Lacy1994}. The black line shows the direction of the profile in Fig.~\ref{fig:grad}.}

\label{fig:RD_fd}
\end{figure*}

\begin{figure*}
\centering
\includegraphics[scale=0.38]{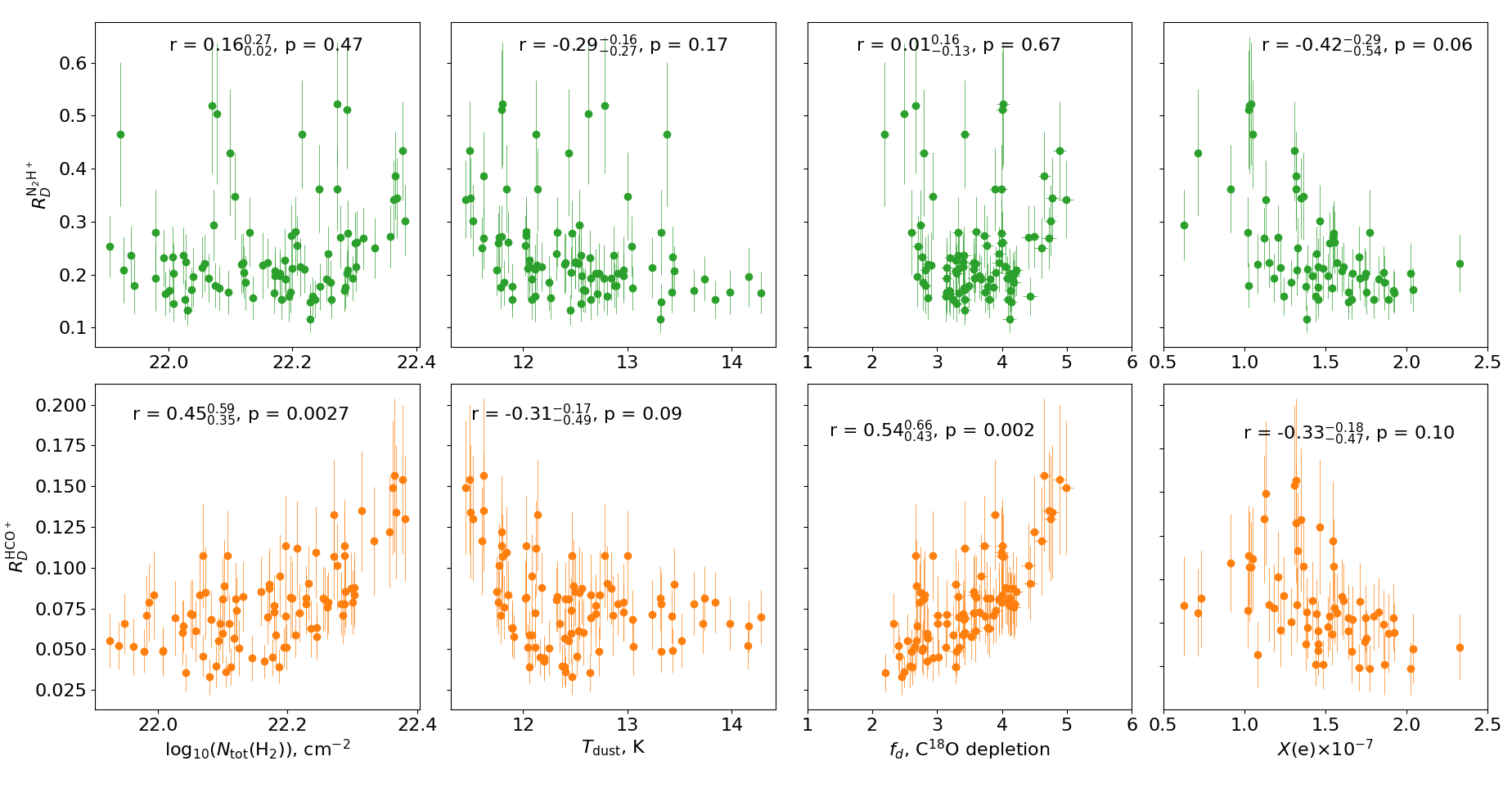}
\caption{Deuterium fraction, $R_D$, as a function of column density of molecular hydrogen, $N(\ce{H2})$, dust temperature, $T_{\rm dust}$, CO depletion factor, $f_d$, and ionization degree, $X(e^-)$. CO depletion factor was calculated with 4.5$\times$10$^{-7}$ from \citet{Lacy1994}.}
\label{fig:R_CO}
\end{figure*}

We divided the derived column density maps of deuterated isotopologues by those of hydrogenated isotopologues pixel by pixel to estimate the deuterium fraction. Figure~\ref{fig:RD_fd} shows the deuterium fraction maps of \ce{N2H^+} (far left) and \ce{HCO^+} (center left). The deuterium fraction of carbon-bearing \ce{HCO^+}, $R_D^{\rm HCO^+}$, which ranges from 0.021$\pm$0.005 to 0.094$\pm$0.021, is by an order of magnitude smaller than that of nitrogen-bearing \ce{N2H^+}, $R_D^{\rm N_2H^+}$, which ranges from 0.11$\pm$0.03 to 0.52$\pm$0.14, like in the dense cores of L1688 \citep{Petrashkevich2024}. The deuterium fraction increases toward the dust peak of the starless core, as was found in other starless cores \citep{Caselli2002,Crapsi2004,Crapsi2007,Pagani2007,Pagani2009b,Chacon-Tanarro+2019,Redaelli2019,Petrashkevich2024}. We estimate how the YSO affects the deuterium fraction on the scale of our beam size, that corresponds to $\simeq$10000~au. The deuterium fraction toward the YSO is lower than that toward the starless core by a factor of $\sim$2 for $R_D^{\rm HCO^+}$ and a factor of 3 for $R_D^{\rm N_2H^+}$ (see Table~\ref{tab:column}). 

\citet{Friesen2013} presented the deuterium fraction $R_D^{\rm N_2H^+}$ toward three positions in B5 (the protostellar core, the starless core, and a position to the north of the protostellar core), marked with black crosses in Fig.~\ref{fig:RD_fd}. Toward these positions, our estimates of deuterium fraction are higher by a factor of 2--7 because we obtained higher $N_{\rm tot}$(\ce{N2D^+}) and lower $N_{\rm tot}$(\ce{N2H^+}) (see Sect.~\ref{sec:col_den}). Our values of deuterium fraction for both species are similar to those of cold dense cores in the L1688 region \citep{Petrashkevich2024}. \citet{Roberts2007} obtained the deuterium fractions $R_D$(HDCO/H$_2$CO) = 0.02--0.11 and $R_D$(D$_2$CO/H$_2$CO)$<0.09$ toward the protostellar core of B5, similar to our $R_D^{\rm HCO^+}$. This is consistent with the idea that carbon-bearing species trace the lower deuteration of core envelopes \citep[see, e.g.,][]{Petrashkevich2024}. Toward the protostellar core in B5, \citet{Hatchell_2003} estimated $R_D$(\ce{NH2D}/\ce{NH3}) to be 0.18, which is consistent with our $R_D^{\rm N_2H^+}$ = $0.15\pm0.04$, and \citet{Imai2018} estimated $R_D$(DNC/HNC) to be 0.05--0.06, similar to our $R_D^{\rm HCO^+}=0.05 \pm 0.01$, as was observed in protostellar cores before \citep[e.g., in HH211,][]{Giers2023}. Our $R_D$(\ce{NH2D}/\ce{NH3}) $=0.36\pm0.20$ toward the starless core is also consistent with the previous findings. 

\begin{figure}
\centering
\includegraphics[scale=0.54]{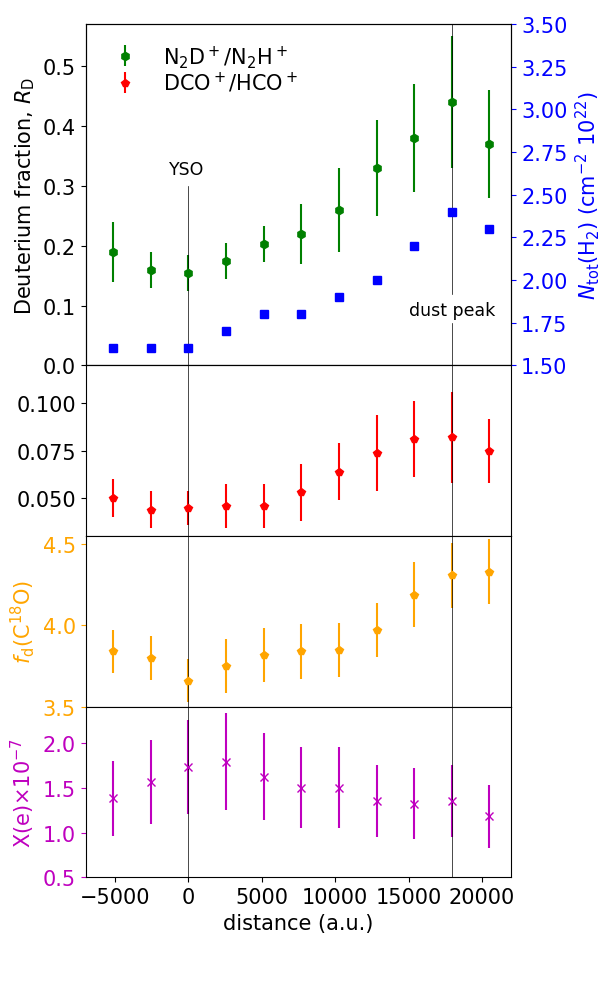}
\caption{The profile of average deuterium fraction, molecular hydrogen column density, CO depletion factor, and ionization degree in the direction from the protostar to the dust peak of the starless core.} 
\label{fig:grad}
\end{figure}

We searched for linear correlations using the Pearson coefficient between the deuterium fraction and the physical parameters of the medium: gas and dust temperature ($T_{\rm gas}$, $T_{\rm dust}$), molecular hydrogen column density ($N_{\rm tot}$(H$_2$)), ionization degree, CO depletion factor, and turbulence, represented by the non-thermal component of the velocity dispersion \citep[$\sigma_{\rm NT}=\sqrt{\sigma^2-k T_k/m_{\rm obs}}$, where $m_{\rm obs}$ is the mass of the observed molecule in a.m.u.; for details, see Sect. 4.2.1 of][]{Petrashkevich2024}. We used the dust temperature and molecular hydrogen column density maps from \citet{Pezzuto2021}, and the gas temperature was calculated using the HFS fit of the ammonia data from \citet{Pineda2021} as described in Sect.~\ref{sec:hfs-fit}. The gas and dust temperature maps are presented in the left and central panels of Fig.~\ref{fig:T_g} (in Appendix A). The lower limit on the ionization degree is given as a sum of the abundances of all observed ions (\ce{N2H^+}, \ce{N2D^+}, \ce{HCO^+}, and \ce{DCO^+}, see the right panel of Fig.~\ref{fig:RD_fd}). We calculated the Pearson coefficient taking into account the 2D distribution (the maps) and estimated the effective number of independent data points to correctly assess confidence intervals. We present the median correlation coefficient ($r$) with 95\% confidence intervals, and the probability ($p$) of getting the $r$ value under the null hypothesis of no correlation ($r=0$), that represents statistical significance. The probability was calculated in 5000 Monte Carlo iterations.

We found stable significant correlations for the deuterium fraction $R_D^{\rm \ce{HCO+}}$ with $N({\rm \ce{H2}})$ ($r=0.45^{0.59}_{0.35}$, $p=0.0027$) and $f_d$ ($r=0.54^{0.66}_{0.43}$, $p=0.0002$) and a weak anticorrelation with $T_{\rm dust}$ ($r=-0.31^{-0.17}_{-0.40}$, $p=0.09$, see the bottom panels of Fig.~\ref{fig:R_CO}). The correlation of $R_D^{\rm HCO^+}$ with $f_d$ looks like increasing CO depletion factor sets a lower boundary on the deuterium fraction (see Fig.~\ref{fig:R_CO}). We did not find significant correlation between $R_D^{\rm \ce{HCO+}}$ and $X(e^-)$ ($r=-0.33_{-0.47}^{-0.18}$, $p=0.10$), $T_{\rm gas}$ ($r=-0.08^{0.04}_{-0.21}$, $p=0.6$), $\sigma_{\rm NT}$ ($r= 0.09^{0.25}_{-0.01}$, $p=0.54$). We also did not find any statistically significant correlations between $R_D^{\rm \ce{N2H+}}$ and the physical parameters (the Pearson coefficients are given in parentheses): with $N(\rm \ce{H2})$ ($r= 0.16^{0.27}_{0.02}$, $p=0.47$), $T_{\rm dust}$ ($r=-0.29^{-0.16}_{-0.38}$, $p= 0.17$), $f_d$ ($r= 0.01^{0.16}_{-0.13}$, $p=0.67$), $T_{\rm gas}$ ($r=-0.04^{0.09}_{-0.12}$, $p=0.9$), and $\sigma_{\rm NT}$ ($r= 0.12^{0.32}_{0.031}$, $p=0.55$). The only correlation for $R_D^{\rm \ce{N2H+}}$ was found with $X(e^-)$: $r=-0.42_{-0.54}^{-0.30}$, $p=0.06$. The spread of data points, the uncertainties of the deuterium fractions, and the small range of the parameters resulted in poor significance of the correlations.   

\section{Discussion}\label{sec:discussion}

\subsection{Canonical CO Abundance}\label{sec:discus_depletion}
One of the objectives of our study was to understand why the starless core in B5 showed relatively high deuterium fraction of \ce{N2H^+} and a low CO depletion factor \citep[$R_D$=0.20$\pm$0.04 and $f_d$=1.43, core 6 in][]{Friesen2013}. Our estimates of both $R_D$=0.4$\pm$0.1 and $f_d$=4.6$\pm$0.1 toward the same position are higher and more consistent with the expectations of strong CO depletion in the region with high deuteration. Our $N$(C$^{18}$O) is 15\% larger than in \citet{Friesen2013}. Both works use the same transition of C$^{18}$O(2--1), and the difference in column density is likely due to beam dilution, since we convolved the C$^{18}$O(2--1) map with the common 33.6$^{\prime\prime}$ beam that is larger than the native 11.8$^{\prime\prime}$ beam. However, our CO depletion value is larger by a factor of 3.2. The biggest contribution to this difference is introduced by the factor of 2.9 larger canonical CO abundance that we use \citep[$X$(C$^{18}$O) = $4.8\times10^{-7}$ of \citet{Wilson1994_HCO} compared to $1.7\times 10^{-7}$ of][]{Frerking1982}. Our $N$(H$_2$) is 40\% higher, possibly because of the data used to estimate the column density of molecular hydrogen -- {\it Herschel} dust continuum emission at 250~$\upmu$m, 350~$\upmu$m, 500~$\upmu$m tracing warmer dust compared to colder 850~$\upmu$m dust from the SCUBA survey used in \citet{Friesen2013}, although usually the probe of colder dust in molecular clouds results in a higher $N$(H$_2$) \citep[e.g.,][]{Howard2019,Pezzuto2021}. The combination of these three factors gave a factor of 3.2 higher $f_d$, and with the native beam $f_d$ would be even higher. A higher $N$(H$_2$) compensated for the difference introduced by the beam dilution. That is, the canonical abundance of CO is the largest contributor to the difference in CO depletion factor. 

While 850~$\upmu$m emission better traces the cold dust and thus the gas of the starless core, {\it Herschel/SPIRE} data better trace the warmer dust and gas of the cloud. It might be not the best tracer of $N$(H$_2$) in dense cores but it traces the gas of the entire cloud and thus is a perfect instrument to estimate the abundance of the undepleted $^{13}$CO and to choose the most suitable canonical abundance of CO, as presented in Sect.~\ref{sec:CO_depl}. Our results on the CO depletion factor in B5  are more consistent with the definition of CO depletion, since at all positions toward this dense clump the value of CO depletion factor is $>1$. 

\subsection{High Deuterium Fraction Overall in B5}

The difficulty in fitting HFS structures to spectra with moderate and low S/N is in the large uncertainty in the spectrum parameters, especially the optical depth. In this case, the estimation of the column densities of species is closely related to the accuracy of the optical depth. There are several methods to treat such spectra, for example, the lines with low optical depth can be considered optically thin \citep[e.g.,][]{Caselli2002,Crapsi2005} or using several transitions (this is the method we used for ammonia, see Sec. \ref{sec:ammonia}). We applied the method described in \citet{Petrashkevich2024}, which utilises the Monte Carlo method to explore the correlation between $T_{\rm ex}$ and $\tau$ and applies the HFS fit with constrained $T_{\rm ex}$ to better estimate $\tau$. As noted in \citet{Petrashkevich2024}, a 1--2~K incorrect $T_{\rm ex}$ contributes to a 10--20\% variation in the column density. The dust temperature difference toward the starless and protostellar cores is $\sim$2~K, and the difference between the excitation temperatures of \ce{N2H^+}(1--0) and \ce{N2D^+}(1--0) is even smaller, $\sim$1~K (see Sec. \ref{sec:hfs-fit}), that is using the same $T_{\rm ex}$ within each map does not significantly affect resulting deuterium fraction.

There are other dense cores that showed similarly high deuterium fractions of \ce{N2H^+} and \ce{NH3} of 0.2--0.7 \citep[L1544, L183, IRAS03282, B1b, HH211, L1688;][]{Caselli2002,Crapsi2007,Pagani2007,Roberts2007,Emprechtinger2009,Punanova2016,Petrashkevich2024}. L1544, L183, and part of the cores in L1688 are isolated from protostellar feedback \citep[e.g.,][]{Caselli2017,Pagani2003,Punanova2016}. The high deuteration is possibly reached there due to the combination of high density, low temperature, low ionisation, and advanced stage of chemical evolution \citep[L1544 is considered to be the prototypical prestellar core at advanced stage of chemical evolution;][]{Caselli2010,Caselli2022,Redaelli2019}, as predicted by chemical models \citep[see, e.g.,][]{Sipila2010,Sipila2013,Sipila2019,Kong2015}. However, the cores IRAS03282, B1b, HH211 \citep{Roberts2007,Emprechtinger2009}, and Oph-F in L1688 \citep{Petrashkevich2024} are protostellar, with embedded Class 0 sources. \citet{Emprechtinger2009} show that deuterium fraction drops significantly as the bolometric luminosity, a measure of protostellar evolution, increases. \citet{Yu1999} and \citet{Zapata2014} showed that B5 IRS1 (IRAS~03445+3242) is possibly in transition from Class 0 to I; this is also supported by the spectral index $\beta=0$ obtained by \citet{Emprechtinger2009}. Probably high deuterium fraction in the original prestellar core and the early evolutionary stage of the YSO explain the still high deuterium fraction of \ce{N2H^+} observed toward the protostellar core. Another possible reason why the deuterium fraction of \ce{N2H^+} is high toward the protostellar core may be the contribution from the starless cores close to B5-IRS1 \citep[B5-Cond2 and B5-Cond3, see][]{Pineda2015} due to beam dilution effect. 

\subsection{Deuterium Fraction across B5}

The integrated intensity maps (see Fig.~\ref{fig:W-maps}) and the line profiles (see Fig.~\ref{fig:spec} in Appendix A) show that the line intensity of the hydrogen-bearing species is higher toward the protostellar core and the line intensity of the deuterated species is higher toward the starless core. This might be an effect of the YSO that heats the surrounding gas, starting the reverse deuterium fractionation \citep[decreased intensity of D-bearing species; e.g.,][]{Caselli2012}, an effect of the higher density toward the starless core on a scale of our beam enhancement of D-bearing species, or due to different stages of chemical evolution of the filaments (both effects). Here we analyse the spatial distribution of the deuterium fraction in relation to CO depletion, the gas density, and the structure of the B5 clump.

Despite the early evolutionary stage, B5 IRS1 affects the surrounding medium, as illustrated by the increased dust temperature (see central panel of Fig. \ref{fig:T_g} in Appendix A), so it probably affects local deuteration too. Figure \ref{fig:grad} shows the change of the deuterium fraction in both tracers along the line passing through the YSO and the peak deuterium fraction $R_D^{\rm N_2H^+}$ associated with the dust peak of the starless core (the line is shown in Fig.~\ref{fig:RD_fd}). 
We averaged three incident pixels along the line (that is adjacent pixels in the direction perpendicular to the line) with a step of a half pixel diagonal (or 8.5$^{\prime\prime}$, equivalent to $\sim$2560~au, no resampling). The deuterium fraction in \ce{N2H^+} increases by a factor of $\sim3$ from 0.15$\pm$0.03 toward IRS1 to 0.44$\pm$0.11 toward the starless core. The deuterium fraction of HCO$^+$ increases by a factor of 2, from $0.04\pm0.01$ to $0.08\pm0.02$. The increase correlates with the molecular hydrogen column density on the right from YSO and does not correlate on the left from YSO in the plot. The deuterium fraction increases in both directions from IRS1, and at the distance of 7500~au from the protostar to the starless core, the increase becomes steeper. Along the profile, both $R_D^{\rm N_2H^+}$ and $R_D^{\rm HCO^+}$ increase with the CO depletion factor that also increases in both directions from IRS1. 

\citet{Schmiedeke2021}, based on interferometric ammonia observations, suggested that the B5 clump consists of two filaments. We estimated the average deuterium fraction in both filaments using the profiles from \citet{Schmiedeke2021}. The average $R_D$ is larger in filament~1 with the starless core than in filament~2 with the YSO in both \ce{N2H^+} (0.28$\pm$0.09 vs 0.17$\pm$0.04) and \ce{HCO^+} (0.06$\pm$0.01 vs 0.04$\pm$0.01, see Table~\ref{tab:RD_Filment}). We assume that the deuterium fraction in the two filaments may be different due to the protostar. The protostar affects the filament gas where it is embedded, including the neighbouring starless cores \citep[B5-Cond2 and B5-Cond3;][]{Pineda2010}, but has much less effect on the starless core in the neighbouring  filament~1. The average central density of filament~1 and filament~2 are the same within the uncertainties, ($1.7\pm0.7$) and ($1.4\pm0.6$) $\times10^6$~cm$^{-3}$, so it could not affect deuterium fraction.

We suggest that the deuterium fraction in the dense cores near the YSO is decreased by a small increase in gas and dust temperatures and by the protostellar feedback. Models predict that deuterium fraction decreases with higher temperature \citep[$>13$~K][]{Kong2015}, and the abundance of \ce{N2D^+} decreases faster than that of \ce{N2H^+} with the temperature \citep{Sipila2018}. In the model, the most noticeable decrease starts after 15~K, most probably due to the increasing ortho-to-para \ce{H2} ratio in the gas, which stays around $\sim3\times10^{-4}$ at 5--15~K and increases by an order of magnitude over 15--20~K and by another order of magnitude over 20--25~K \citep{Kong2015}. The more energetic ortho-\ce{H2} drives the proton exchange reaction backward, decreasing abundance of \ce{H2D^+} \citep[e.g.,][]{Sipila2013}. 

The negative correlation between $R_D^{\ce{HCO+}}$ and $T_{\rm dust}$ also indicates the effect of the ortho-to-para \ce{H2} ratio: it is expected to decrease with increasing volume density (and thus decreasing $T_{\rm dust}$), which drives deuteration \citep{Sipila2013}. 
If the $R_D^{\ce{HCO+}}$ -- $T_{\rm dust}$ anticorrelation indeed reflects the increase in the ortho-to-para \ce{H2} ratio, we should see a similar anticorrelation between $T_{\rm dust}$ and deuterated ions, while hydrogenated ions should not be affected. To test this prediction, we calculated the Pearson coefficients (see Sect.~\ref{sec:rd}) for the correlations of $X(\ce{N2H+})$ and $X(\ce{DCO+})$ (the ions having the largest map size) with $T_{\rm dust}$ and $T_{\rm gas}$. $X(\ce{N2H+})$ does not correlate with ether $T_{\rm dust}$ ($r=0.075_{0.034}^{0.111}$, $p=0.56$) and $T_{\rm gas}$ ($r=-0.033_{-0.070}^{-0.0006}$, $p=0.80$). $X(\ce{DCO+})$ shows an insignificant anticorrelation with $T_{\rm dust}$ ($r=-0.17_{-0.22}^{-0.12}$, $p=0.20$) and a significant anticorrelation with $T_{\rm gas}$ ($r=-0.26_{-0.30}^{-0.21}$, $p=0.05$). This test results do not prove a strong relation of the ortho-to-para \ce{H2} ratio and $R_D$; a data set with a wider range of $T_{\rm dust}$ and $R_D$ and smaller uncertainties would clarify this question.

\begin{table}
\caption{Average deuterium fraction toward filaments with the YSO and the starless core. The pixels for the average were selected following the profiles from \citet{Schmiedeke2021}. The number of pixels used for the average value is given in parentheses.}\label{tab:RD_Filment}
\begin{tabular}{lcc}
\hline\hline
Deuterium fraction &  fil2: YSO & fil1: starless core \\ \hline
$R_D^{\rm N_2H^+}$ &  0.17$\pm$0.04 (38) &  0.28$\pm$0.09 (44) \\
$R_D^{\rm HCO^+}$ & 0.04$\pm$0.01 (25) &   0.06$\pm$0.01 (30)\\ \hline
\end{tabular}
\end{table}

\section{Conclusions}\label{sec:conclusion}

We presented the study of deuterium fractionation in B5 in the Perseus molecular cloud. We estimated column densities, deuterium fractions, and CO depletion factor through observation maps of the \ce{N2H^+}(1--0), N$_2$D$^+$(1--0), H$^{13}$CO$^+$(1--0), DCO$^+$(2--1), and \ce{C^{18}O}(2--1) lines. We chose the most suitable canonical CO abundances by calculating the CO depletion factor based on the $^{13}$CO(1--0) observations from the COMPLETE survey \citep{Ridge2006} for the entire Perseus molecular cloud. We analysed how a young protostar affects the environment including the deuterium fraction, the ionization degree and the CO depletion and estimated the variation of these environmental factors along the profile from the protostar to the starless core. Our conclusions are as follows.

\begin{enumerate}

\item We presented the first maps of deuterium fraction $R_D^{\rm N_2H^+}$ and $R_D^{\rm HCO^+}$ and CO depletion factor toward B5. The deuterium fractions are in the ranges $R_D^{\rm N_2H^+}$ = 0.11--0.52 and $R_D^{\rm HCO^+}$ = 0.021--0.094; they are comparable to those previously observed in other cold dense cores in Perseus and other low-mass star-forming regions. The deuterium fraction in nitrogen-bearing \ce{N2H^+} is by a factor of 5 higher than that in carbon-bearing \ce{HCO^+}, similar to findings in other regions of star formation \citep[e.g., L1544, L1688][]{Caselli2002,Petrashkevich2024} and confirms that the deuterium fraction increases with density in starless cores. The decreasing deuterium fraction is observed toward the protostar.

\item We presented the distribution of the CO depletion factor toward the Perseus molecular cloud, the L1688 star-forming region, and $f_d$ map of B5. By analysing the distributions of different star-forming regions, the best canonical CO abundance $X$(CO)$_{\rm can}$ = 2.69$\times$10$^{-4}$ from \citet{Lacy1994} was selected as most appropriate to calculate the CO depletion factor in the Perseus molecular cloud. The values obtained for the CO depletion factor in B5 are higher than those obtained earlier in \citet{Friesen2013}. The new values of the CO depletion factor are in line with theoretical expectations, given the high degree of deuterium fractionation observed in the B5 region.

\item For the first time we studied how deuterium fraction changes around a low-mass protostar next to a starless core. We plotted the average profile of the deuterium fraction from the protostar to the dust peak of the starless core. The deuterium fraction $R_D^{\rm N_2H^+}$ shows a gradient with a minimum toward the protostar and a maximum toward the dust peak of the starless core, consistent with deuterium fractionation theory \citep{Dalgarno1984}. 

\item We found statistically significant correlations between deuterium fraction $R_D^{\ce{HCO+}}$ and molecular hydrogen column density, $N(\ce{H2})$ ($r=0.45^{0.59}_{0.35}$, $p=0.0027$) and CO depletion factor, $f_d$ ($r=0.54^{0.66}_{0.43}$, $p=0.0002$). We found weak anticorrelations between $R_D^{\ce{HCO+}}$ and dust temperature, $T_{\rm dust}$ ($r=-0.31^{-0.17}_{-0.40}$, $p=0.09$) and between $R_D^{\ce{N2H+}}$ and ionization degree, $X(e^-)$ ($r=-0.42_{-0.54}^{-0.30}$, $p=0.06$). These positive and negative correlations are consistent with model predictions. We did not find statistically significant correlation between the deuterium fractions and gas temperature, $T_{\rm gas}$, and level of turbulence (represented by $\sigma_{\rm NT}$), due to the small range of the parameters and their relatively large uncertainties. 

\item We explored the parameter space $\tau-T_{\rm ex}$ to find and constrain the excitation temperature to better estimate the optical depth of the lines \citep[as was done in][]{Petrashkevich2024}. This method allows to estimate the column densities of weaker deuterated species more accurately and obtain correct deuterium fractions. 

\end{enumerate}

The obtained distributions of the deuterium fraction and the CO depletion factor can be used to benchmark astrochemical models. CO depletion factor was utilized to constrain chemical models of prestellar cores before \citep[e.g.,][]{Vasyunin2017,Borshcheva2025}.  The unique structure of B5 allows us to study the deuteration process around a protostar and in a starless core within one object. Further study of deuterium fractionation in B5 requires observations with a higher, interferometric, angular resolution.






\begin{acknowledgments}
The authors thank the anonymous referee for valuable comments that helped to improve the manuscript. IRAM is supported by INSU/CNRS (France), MPG (Germany) and IGN (Spain). The work of A.P. and A.V is supported by the RSF project 23-12-00315 (Sections \ref{sec:CO_depl} and \ref{sec:discus_depletion}). I.P. acknowledges the financial support of the Ministry of Science and Education of Russia, the FEUZ-2025-0003 project. 
\end{acknowledgments}

\facilities{IRAM 30~m, GBT, FCRAO, HST(SPIRE)}

\software{\textsc{pyspeckit} \citep{Ginsburg2011,Ginsburg2022}, \textsc{astropy} \citep{Astropy2013,Astropy2018,Astropy2022}, \textsc{gildas} \url{https://www.iram.fr/IRAMFR/GILDAS/}.  
          }


\appendix

\section{Maps, Spectra, and Plots}

\begin{figure}
\centering
\includegraphics[scale=0.5]{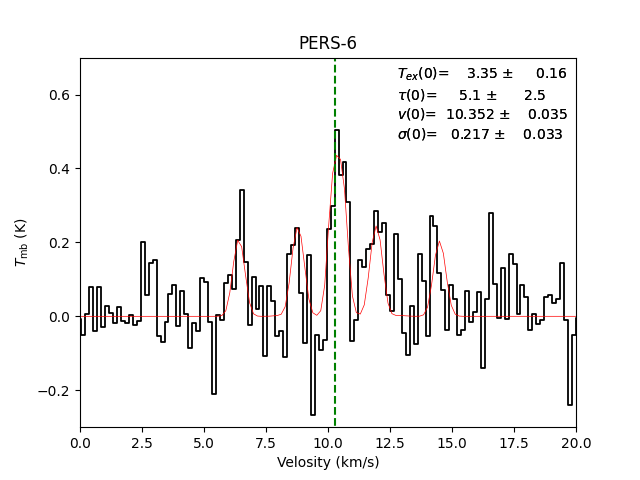}
\caption{The spectrum (black) and the fit (red) of the \ce{p-NH2D}(1,1) line toward the protostellar core. The fit parameters, obtained with \textsc{pyspeckit}, were used to calculate $N$(\ce{p-NH2D}) are given in the top right corner.}
\label{fig:NH2D}
\end{figure}

\begin{figure*}
\centering
\includegraphics[scale=0.29]{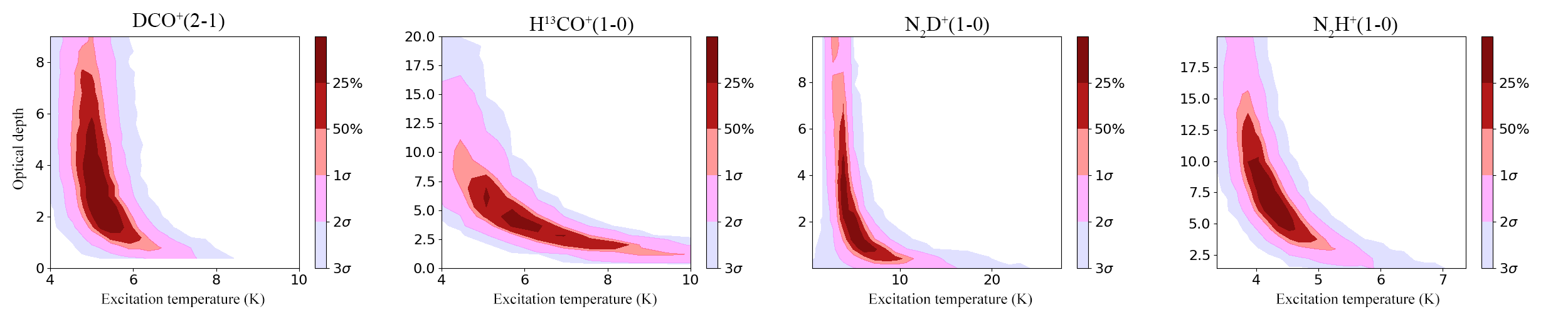}
\caption{Parameter space exploration of the optical depth and the excitation temperature of the observed lines performed with the Monte Carlo method. Abscissa and ordinate show the excitation temperature and the optical depth, respectively. The color scale shows the percentiles corresponding to the 1-, 2- and 3-$\sigma$ confidence intervals of the normal distribution (68.27, 95.45 and 99.73\%).}
\label{fig:zone1}
\end{figure*}

\begin{figure*}
\centering
\includegraphics[scale=0.55]{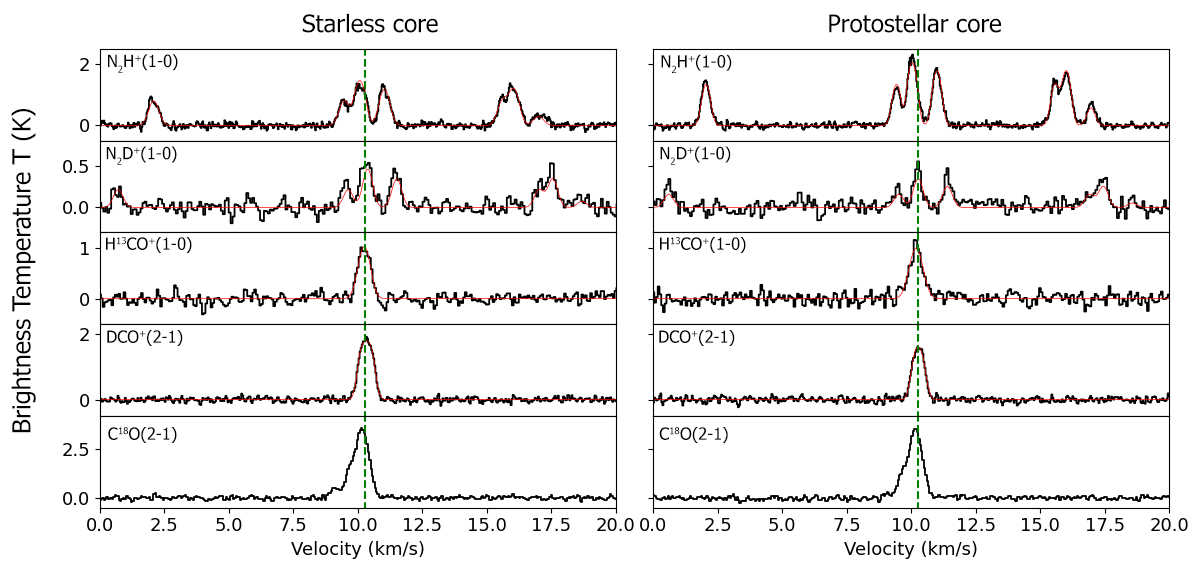}
\caption{The spectra (black) and the fits (red) of the observed lines toward the dust peak of starless core (left) and the protostellar core (right). The dashed green vertical lines show the $V_{\rm LSR}$ toward each position, averaged over all lines.} 
\label{fig:spec}
\end{figure*}

\begin{figure*}
\centering
\includegraphics[scale=0.27]{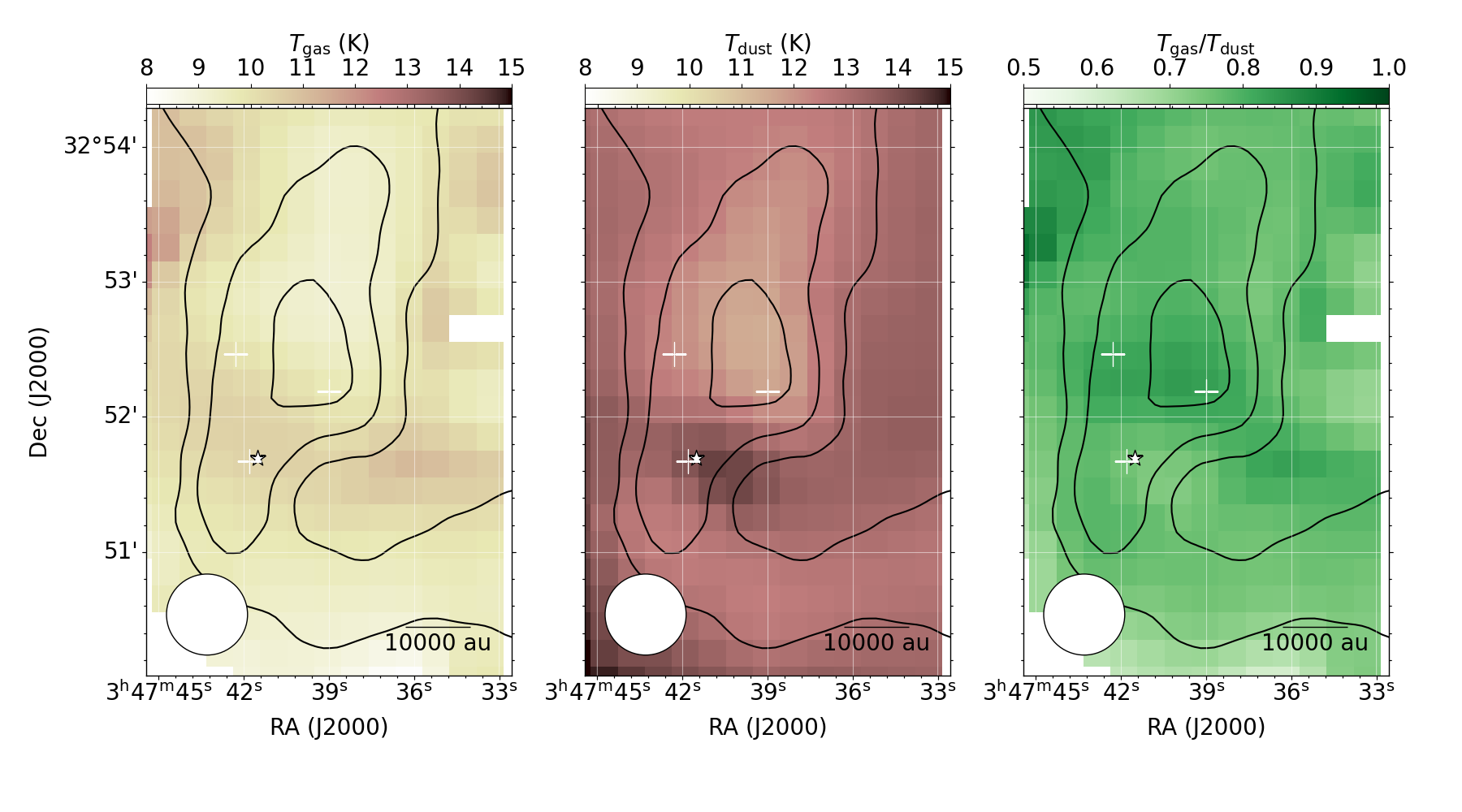}
\caption{{\it Left:} the gas temperature obtained with the \ce{NH3} (1,1) and (2,2) line analysis. {\it Center:} the dust temperature was taken from  Herschel Gould Belt Survey Archive \citep{Pezzuto2021}. {\it Right:} the ratio of the gas and dust temperatures. The black contour shows the column density of molecular hydrogen. The first contour starts at 0.8$\times$10$^{22}$~cm$^{-2}$ with a contour step of 0.5$\times$10$^{22}$~cm$^{-2}$. The star shows the position of the YSO \citep{Yu1999}. The beam size is shown in the bottom left corner of each map. The white crosses show the positions observed by \citet{Friesen2013}.} 
\label{fig:T_g}
\end{figure*}

\begin{figure*}
\centering
\includegraphics[scale=0.5]{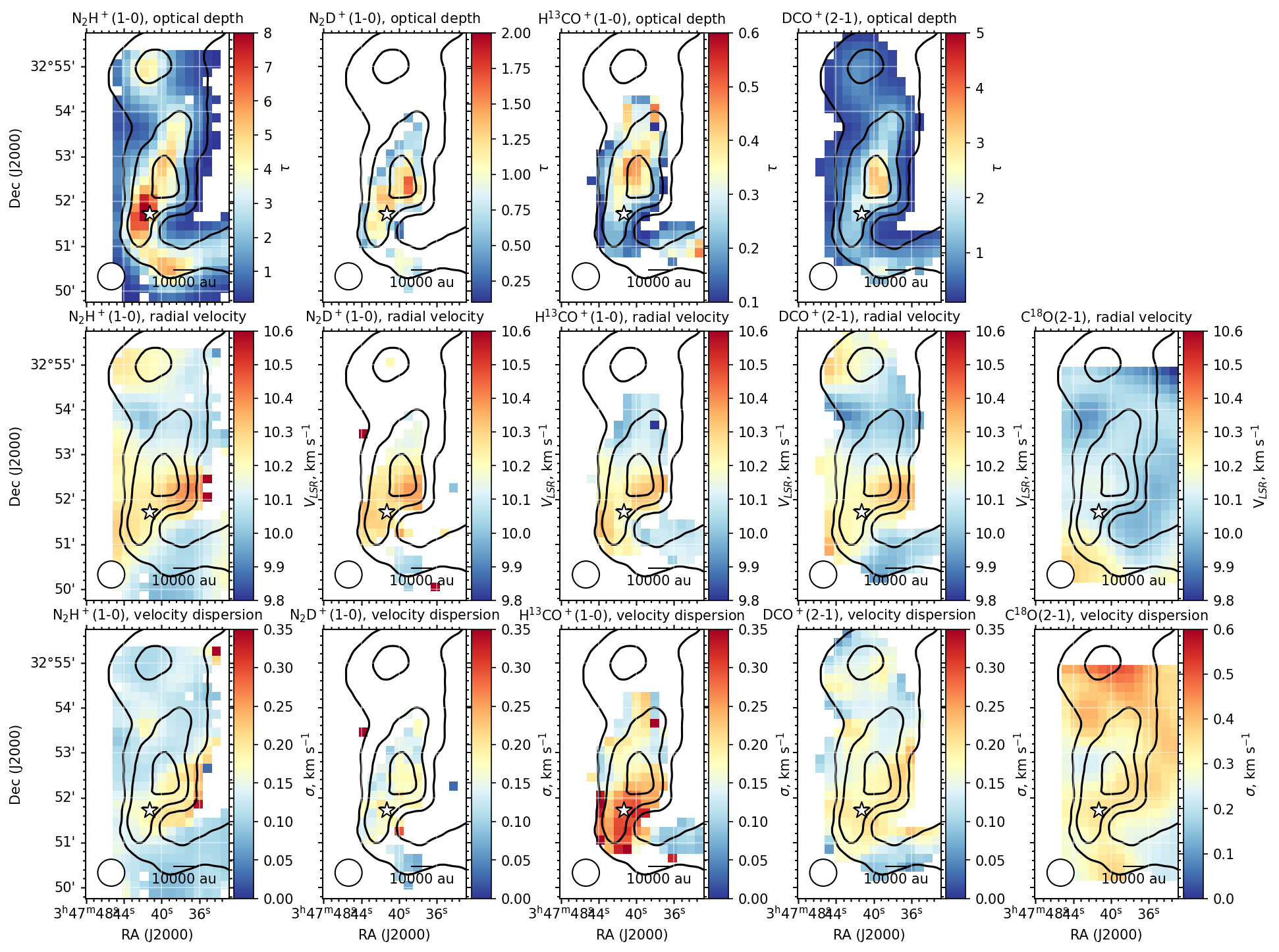}
\caption{Parameter maps for all species. {\it Top:} optical depth ($\tau$); {\it center:} radial velocity (V$_{\rm LSR}$); {\it bottom:} velocity dispersion {$\sigma$}. The black contour shows the column density of molecular hydrogen. The first contour starts at 0.8$\times$10$^{22}$~cm$^{-2}$ with a contour step of 0.5$\times$10$^{22}$~cm$^{-2}$. The star shows the position of the YSO \citep{Yu1999}. The beam size is shown in the bottom left corner of each map. In the ``outlying'' pixels, the fit fitted noise features instead of the lines because either the lines are very narrow or the S/N is low. These pixels do not appear in the maps of deuterium fraction and do not affect our analysis.}
\label{fig:tau}
\end{figure*}

\begin{figure*}
\centering
\includegraphics[scale=0.25]{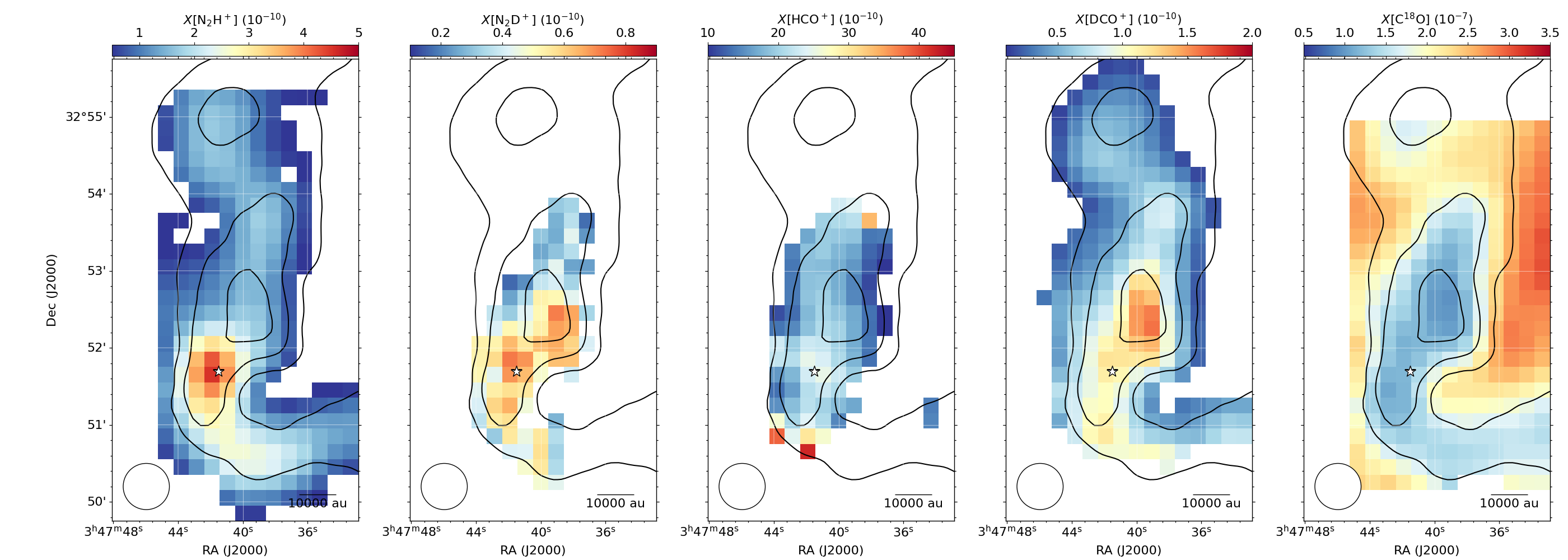}
\caption{The abundance maps of N$_2$H$^+$, N$_2$D$^+$, DCO$^+$, HCO$^+$, and C$^{18}$O toward B5. The black contour shows the column density of molecular hydrogen (the first contour starts at 0.8$\times$10$^{22}$~cm$^{-2}$ with a contour step of 0.5$\times$10$^{22}$~cm$^{-2}$). The star shows the position of the YSO \citep{Yu1999}. The beam size is shown in the bottom left corner of each map.}
\label{fig:X}
\end{figure*}

Here we present the spectrum of \ce{p-NH2D}(1,1) toward the starless core (Fig.~\ref{fig:NH2D}); the parameter space exploration of the optical depth and the excitation temperature (Fig.~\ref{fig:zone1}); spectra toward the cold dense core and YSO (Fig.~\ref{fig:spec}); gas and dust temperature (Fig.~\ref{fig:T_g}); line parameter maps (Fig.~\ref{fig:tau}) of $\tau$, $V_{\rm LSR}$ and $\sigma$; and abundance maps (Fig.~\ref{fig:X}). 

\bibliography{B5_literature}{}
\bibliographystyle{aasjournalv7}



\end{document}